\documentclass[%
 reprint,
 aip,
 amsmath,amssymb,
floatfix,
]{revtex4-1}
\usepackage{xcolor}
\usepackage{natbib}
\usepackage{blindtext}
\usepackage{float}
\usepackage{graphicx}
\usepackage{dcolumn}
\usepackage{bm}
\usepackage{soul}
\usepackage{url}
 \usepackage{hyperref}     
 
\begin{document}
\preprint{APS/123-QED}


\title{
First-principles based study of magnetic states and high-pressure enthalpy landscape of manganese sulfide polymorphs}
\author{Artem Chmeruk}
 \email{chmeruk@gfz-potsdam.de}
\author{Maribel N\'u\~nez-Valdez}
\affiliation{Deutsches GeoForschungsZentrum GFZ, Telegrafenberg, 14473 Potsdam, Germany}
\affiliation{Institut f\"ur Geowissenschaften, Goethe-Universit\"at Frankfurt, D-60438 Frankfurt am Main, Germany}
\date{\today}%
             %

\begin{abstract}
Using first-principles calculations in combination with special quasirandom structure and occupation control matrix methods, we study the magnetic ordering and the effect of pressure on manganese sulfide polymorphs. At ambient conditions, MnS is commonly observed in paramagnetic rock-salt structure, but as temperature decreases at constant pressure it becomes antiferromagnetic. On the other hand, at room temperature MnS has shown to undergo structural transformations as pressure increases. Here, we show that our approach involving the ordering/disordering of the local magnetic moments in addition to the explicit control of the localization of the Mn $d$-electrons produces energy band gaps and local magnetic moments in excellent agreement with those observed experimentally, particularly for paramagnetic MnS. Finally, we focus on how MnS evolves under pressure and from its enthalpy landscape we identify at about 21~GPa, the structural transformation from rock-salt to orthorhombic MnP-type. This structural transformation resembles closely experimental results in which a new stable but unidentified MnS phase was previously reported. 
\end{abstract}
\maketitle
\section{\label{sec:level1}INTRODUCTION}
Manganese sulfide has had a renew draw of attention due to a number of its possible applications. For example, MnS, similarly to Cd and Zn chalcogenides, is known to form part of diluted magnetic semiconductors that possess attractive magneto-optical properties \cite{tappero1997electronic}. Recently, various nanostructured polymorphs of MnS have attracted wide interest as they have shown a combination of electrochemical properties that make them promising materials for applications in Li-ion and Li-S batteries as well as in supercapacitors \cite{he2014stable,zhang2004mineral,zhang2012unusual,tang2015morphology}. 

At room temperature ($T_R$) and ambient pressure (0~GPa), MnS is a paramagnetic (PM) insulator and it can be found crystallized into face-centered cubic rock-salt (RS) or hexagonal wurtzite (WZ) structures, with the former being the most common in nature known as the mineral alabandite. 
RS-MnS, below its N\'eel temperature ($T_{N} = 150$~K), exhibits antiferromagnetic (AFM) ordering of the second kind (AFM-II), i.e., sheets of ferromagnetically coupled magnetic moments stacked antiferromagnetically in the [111] direction \cite{hastings1956magnetic,linesMnS1966antiferromagnetism}. This particular magnetic configuration in RS-MnS leads to a slight distortion from the ideal cubic structure towards a trigonal symmetry, making RS-MnS a magnetostrictive material \cite{morosin1970striction}. On the other hand, WZ-MnS has a $T_N$ of 80~K \cite{danielian1961exchange} and displays AFM ordering of the third kind (AFM-III) in which two-thirds of the nearest neighbors are antiparallel and one-third parallel while next-nearest neighbors are arranged in the opposite way \cite{hastings1956magnetic}.

As a transition metal (TM) compound, MnS is also a source of active research among theorists and modelers in materials science. The electronic structure of TM systems is notoriously challenging to describe and often requires resorting to specialized theories that go beyond the single-particle description.
Thus, MnS, as its oxide counterpart \cite{zunger2018polymorphous}, is a fruitful and compelling TM system that can be used to test whether or not a single-particle theory such as density functional theory (DFT) \cite{kohn1,kohn2} can correctly determine the system's ground-state and other physical properties. Several works have been carried out in that regard using various implementations of DFT \cite{oguchi1983band,raybaud1997ab,raybaud1997abc, tappero1997electronic,tappero1998comparative,hobbs1999magnetism,rohrbach2003electronic}. Although, some studies correctly predicted the AFM-II configuration as the ground-state of RS-MnS, they showed some discrepancy between their predicted  local magnetic moment ($m_{loc}$) values of 4.92~$\mu$B \cite{tappero1997electronic} and 4.082 ~$\mu$B \cite{hobbs1999magnetism}, and the neutron diffraction experimental measurement of $m_{loc} = 4.54$~$\mu$B at $4.2$~K \cite{fender1968covalency}. On the other hand, in order to study the PM phase of MnS, simulations fall into two categories: I) MnS is modeled as a \emph{non-magnetic} system,  i.e., $m_{loc} = 0$; or II) MnS is modeled with finite but {\it randomly oriented} local magnetic moments. Investigations employing the non-magnetic approach  have the shortcoming that within the single-particle band theory, MnS is always expected to be metallic \cite{raybaud1997abc,hobbs1999magnetism}, since the electron density is uniformly distributed over the orbital states, thus, leading to partially unfilled $d$-bands.
The second approach, more compatible with the actual behavior of the Mn local magnetic moments, makes use of the one-electron energy spectrum determined from multiple-scattering theory, namely the Korringa-Kohn-Rostoker coherent potential approximation (KKR-CPA) method  \cite{oguchi1983band}. Within this picture, non-collinear paramagnetism is achieved when the electron is treated as moving through a medium of completely randomly oriented magnetic moments, then the energy spectrum is obtained by the scattering events that the electron experiences. This scheme, usually called \emph{disordered local moments} (DLM), is inherently better for modeling PM phases than the \emph{non-magnetic} approach described above, as it considers an assembly of finite magnetic moments, which corresponds to the experimentally observed PM state of MnS.  
However, as the CPA effective potential is constructed in such a way as to represent an {\it average effect} of an alloy of randomly oriented moments, the possibility for individual magnetic moments to develop their own local magnetic environments is excluded. Hence, this factor would constitute a critical drawback in the modeling of PM MnS.
An alternative way to modeling magnetic disorder, and the one employed in this work, is the so-called method of \emph{special quasirandom structures} (SQS) \cite{zunger1990special}. Although collinear by construction, it offers an upper hand over the CPA in that, a SQS is built to represent an \emph{average quantity}, in this case the net magnetization. This subtle difference 
in the treatment of finite magnetic moments with respect to the CPA has been shown to be a source of gapping in TM oxides and perovskites \cite{zunger2018polymorphous,zunger2019origin}, and we expect it to play an essential role in TM sulphides as well. For a more detailed discussion on the difference between DLM and SQS techniques, see e.g. \cite{zunger2018polymorphous} and references therein. 

Besides the significance of MnS at ambient conditions, there have been experimental and theoretical attempts to study the effect of pressure on it.
Using X-ray diffraction and diamond-anvil-cell (DAC) experiments, a phase transition from RS (B1) to orthorhombic GeS-type (B16) structure was reported at $\sim$7.2~GPa \cite{kraft1988high}. However, a subsequent independent DAC experiment up to $\sim$21~GPa did not confirm it \cite{mccammon1991static}, while another one also did not observe the B1$\rightarrow$B16 transformation and instead found a transition from RS to an {\it unknown} phase at about 26~GPa that remained stable up to at least 46~GPa  \cite{sweeney1993compression}.  More recent DAC experiments reported the fabrication of a MnS quenchable high-pressure nanostructure, which was  indexed as orthorhombic MnP-type (B31) \cite{b31}. This B1$\rightarrow$B31 transformation was observed at the transition pressure ($P_T$) of approximately 22~GPa, which is in good agreement with the previous findings in bulk MnS, thus shedding light on the nature of the unknown structure. However, the question whether or not there is another structural transition above 46~GPa still remained unanswered. Finally, static DFT simulations of the B1- and B31-MnS phases under pressure suggested that the B31 phase was preferable at all pressures, including ambient conditions \cite{b31}. Though, this DFT study did not specify what magnetic configuration was considered or under what scheme (non-magnetic or disorder PM) the MnS phases were simulated.

Therefore, the purpose of our investigations is twofold. Firstly, we extend the work of \cite{zunger2018polymorphous,zunger2019origin} for TM oxides and perovskites to the case of TM sulphides. Our goal here is to show that without resorting to specialized theories of highly correlated materials beyond on site corrections (DFT+$U$), the insulating PM phase of MnS can be achieved by introducing magnetic disorder using SQS's  and occupation matrix control methodologies to handle localization of Mn $d$-states. In this way, we are able to produce comparable physical properties such as lattice parameter, energy gap, and local magnetic moment to those observed experimentally. Secondly, using different PM MnS polymorphs' ground states as candidates to high-pressure phases, we calculate their enthalpies as a function of pressure in order to pinpoint possible structural transformations of RS-MnS. Here, we emphasize  that the PM MnS polymorphs' structures are constructed and optimized following the SQS formalism previously described.   Thus, in Section II we define the various methods and theoretical considerations used in our work, then we present our results in Section III-A on the AFM phases of MnS in order to compare to previous studies employing different approaches and to available experimental data. Similarly, in Section III-B we report our findings for selected PM MnS polymorphs, and then, in Section III-C we analyze the high-pressure enthalpy landscape of these polymorphs to evaluate the feasibility of structural transformations as a function of pressure. Lastly, we present our summary and conclusions in Section IV. 
\begin{figure*}
\includegraphics[width=0.8\linewidth]{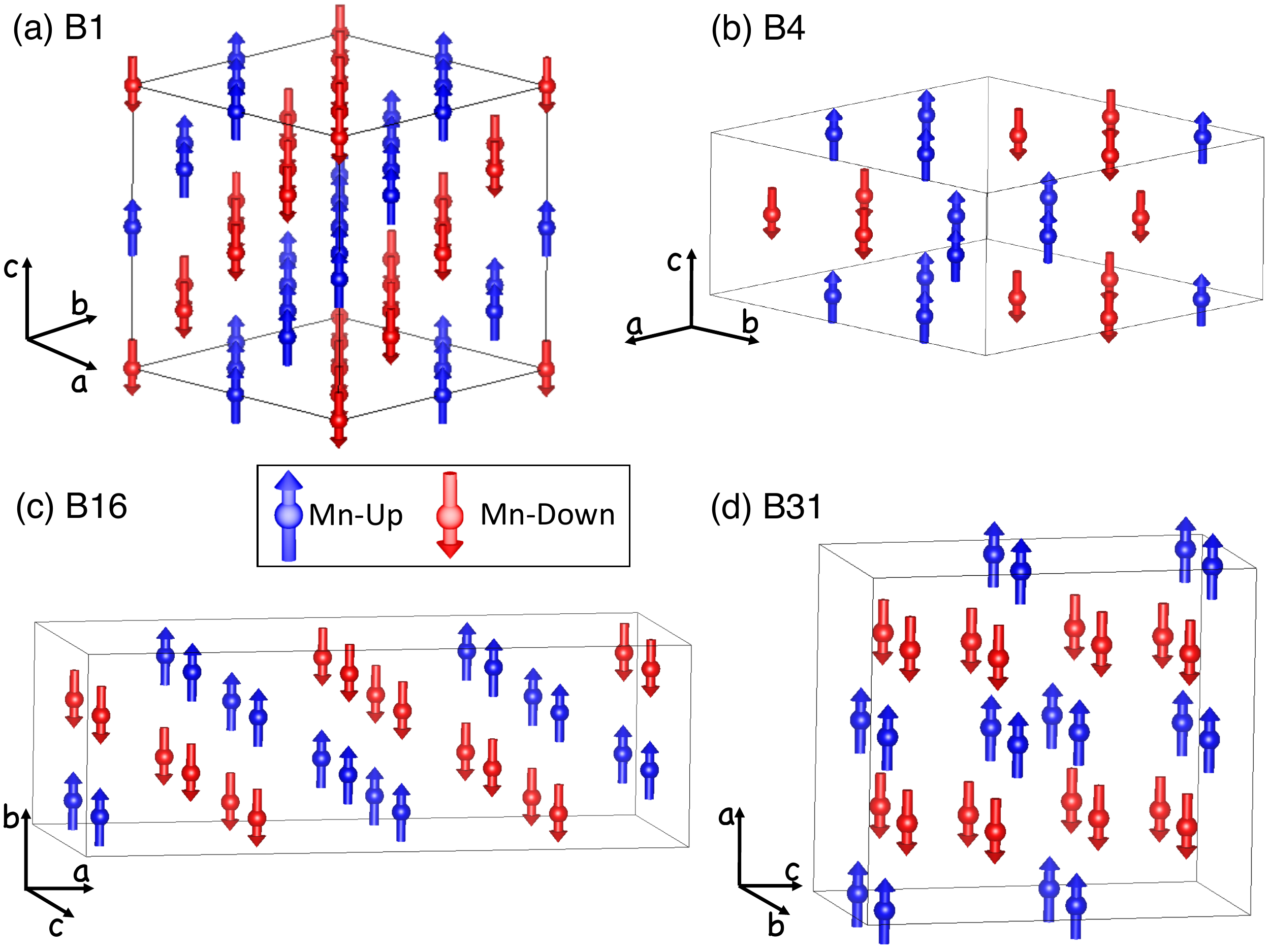}
\caption{\label{afmpolys}AFM arrangements in the MnS polymorphs investigated in this work. (a) B1 with AFM-II ordering and (b) B4 with AFM-III ordering were observed experimentally \cite{hastings1956magnetic};  AFM configurations for the (c) B16 and (d) B31 polymorphs were generated in this study for comparison. S atoms are not shown for simplification. Structures are visualized using VESTA \cite{vesta}}. 
\end{figure*}
\begin{figure*}
\includegraphics[width=0.8\linewidth]{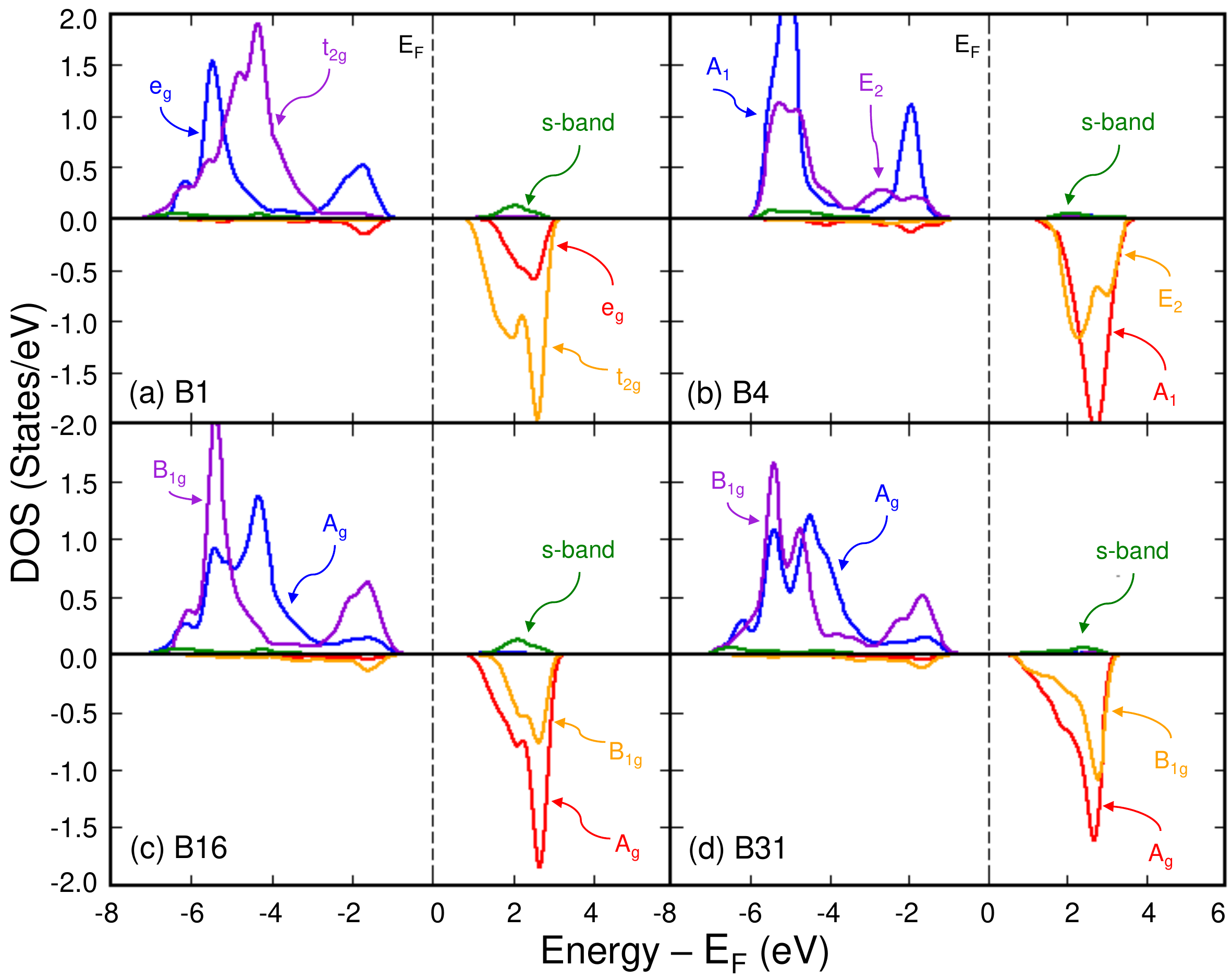}
\caption{\label{afmpdos}DFT$+U$ partial density of states (DOS) for a Mn ion projected on the majority spin (positive) and minority spin (negative) channels with $U=3$~eV within AFM configurations for (a) B1, (b) B4, (c) B16 and (d) B31 MnS polymorphs. The Fermi energy ($E_F$) level is represented by the dashed line at zero. The characters of the $d$-states are interpreted through group theory analysis of the corresponding crystal-field surroundings before the onset of the AFM ordering. This is done for the purpose of comparison with the PM case. AFM ordering lowers the symmetry as follows: in B1 $O_h \rightarrow D_{3d}$, in B4 $C_{6v} \rightarrow C_{1h}$, in B16 $D_{2h} \rightarrow C_{2h}$, and in B31 $D_{2h} \rightarrow C_{2h}$.}
\end{figure*}
\section{\label{sec:level1}COMPUTATIONAL METHODS}
All our calculations were performed within DFT+$U$, using the projector-augmented plane wave method (PAW) \cite{blochl} as implemented in the VASP (version 5.4.4) code \cite{kresse1996efficiency,kresse1996efficient}. The valence configurations were 4\emph{s$^1$}3\emph{d$^6$} for Mn, and 3\emph{s$^2$}3\emph{p$^4$} for S, respectively. The exchange-correlation (XC) term in the effective Kohn-Sham potential was approximated according to  the Perdew-Burke-Ernzerhof parameterization for solids (PBEsol)  of the generalized gradient approximation (GGA) \cite{PBEsol}. To treat the Coulomb repulsion of the Mn $d$-electrons, we added the Hubbard-$U$ correction \cite{AnisimovU} within the  rotationally invariant Dudarev prescription \cite{uterm}:
\begin{equation}
E_{U}= \frac{U}{2}\sum_{I,\sigma}\sum_{i}\lambda^{I,\sigma}_{i}(1-\lambda^{I,\sigma}_{i}). \label{eqU}
\end{equation}
As it is well known, $E_{U}$ represents a penalty energy proportional to $U$ (here $U$ represents the effective difference between the on-site Coulomb and exchange interactions) for atom $I$ and spin channel $\sigma$. 
$\lambda^{I,\sigma}_{i}$ are the eigenvalues (with values between 0 and 1) of the occupation matrix (OM)  $n^{I,\sigma}_{m,m^{\prime}}$  for an orthogonal set of localized orbitals $i$, which, in general, are the linear combinations of the atomic $d$-orbitals $m$. For all our simulations we chose  $U=3$~eV,  as it was already tested for the AFM B1-MnS phase at 0~GPa \cite{rohrbach2003electronic}.
We also assessed the local density approximation (LDA) \cite{ceperley1980ground} and the standard PBE \cite{perdew1996generalized} XC functionals +$U$, but we found that PBEsol+$U$ produced more accurate structural parameters when compared to available experimental values for both AFM and PM phases of MnS. Therefore, we only present results within this scheme in the remaining sections.
Integration in the Brillouin zone was done on a $\Gamma$-centered grid of uniformly distributed k-points with a spacing of $2\pi\times0.3$ \AA$^{-1}$. The selected plane-wave kinetic energy cutoff was 500~eV and convergence of our structural optimizations was assumed when the total energy changes were less than $10^{-8}$~eV and the forces on each atom smaller than $10^{-3}$~eV/\AA.
%
To simulate magnetic ordering (AFM below $T_N$) and disordering (in PM phases at $T_R$) large enough supercells have to be built in order to accomodate the appropriate AFM orderings as well as to allow for multiple relaxation patterns. These two features have been shown to lead to gapping in TM oxides and perovskites \cite{zunger2018polymorphous, zunger2019origin}. Overall, we considered four MnS polymorphs, namely, RS (B1, $Fm\overline{3}m$), wurtzite (WZ or B4, $P6_3mc$), GeS-type (B16, $Pnma$) and MnP-type (B31, $Pnma$). 

To model the structures below $T_N$, we imposed the 
AFM-II and AFM-III orderings observed experimentally in the B1 and B4 polymorphs, respectively \cite{hastings1956magnetic}. For the B16- and B31-MnS structures there is no experimental or computational data on their precise magnetic ordering, except for one report stating that the B31 polymorph is PM above 5~K \cite{b31}. Consequently, we tested several possible AFM arrangements and adopted the lowest energy configurations found. (All AFM input structures are provided in the Supplemental Material \cite{SM}.) We optimized all MnS polymorphs using 64-atom $2 \times 2 \times 2$ supercells, except for B4-MnS, for which a 36-atom $3 \times 3 \times 1$ supercell was sufficient to realize the experimentally observed AFM-III ordering. 

On the other hand, to model the PM MnS polymorphs, we constructed SQS supercells using the  Alloy Theoretic Automated Toolkit (ATAT) software package \cite{van2002alloy}. Under this scheme, the PM state is created as a disordered alloy of {\it up} $\uparrow$ and {\it down} $\downarrow$ moments located at different sites. As the construction of a SQS is based on the computation of the correlation function between the species that constitute the alloy ($\uparrow$ and $\downarrow$ moments in this case), the size of the supercell (number of atoms) used is vitally important to obtain magnetic configurations whose components are not spatially correlated among themselves.  The SQS {\it degree of randomness} is  improved by the number of atomic figures (pairs, triplets, quadruplets, etc.) included in the calculation of the correlation function and by the interaction distance between the atoms in a given figure. 
For example, one could start by only considering nearest neighbors for atomic pairs, then gradually add more pairs (increasing the interaction distance), and/or higher order figures into consideration. The larger the interaction distance and the more figures are considered, the larger the supercell becomes to achieve {\it total randomness}, and the SQS generation quickly becomes computationally demanding. Because the creation of a SQS is purely configurational, in order to save resources, for this study we used a previously produced SQS supercell for the B1 structure of TM oxides \cite{zunger2018polymorphous} with 64 atoms. For the B16- and B31-MnS polymorphs, pairs and triplets were included to obtain 64-atom $2 \times 2 \times 2$ SQS's supercells. Similarly, for the B4 structure, we proved 2$\times$2$\times$2, 3$\times$3$\times$2 and 4$\times$4$\times$1 supercells with 32, 72 and 64 atoms, respectively, resulting in ground state energy differences among them of less than $10^{-3}$~eV per formula unit (f.u.), thus, we took the 32-atom supercell. (All constructed PM structures are given in the Supplemental Material \cite{SM}.) 

Finally, under pressure ($P$) we could not converge some PM MnS structures to stable ground states. In order to deal with this obstacle, it has been noticed that non-integer occupancies in Eq.  \ref{eqU} may lead to local minima and not allow a system to achieve its true ground state. Thus, by {\it controlling} orbital filling explicitly in  $n^{I,\sigma}_{m,m^{\prime}}$, one can {\it help} a non-converging trapped system circumvent a metastable OM ill-setup. This method was shown to find stable states, previously unaccessible, for $d$ and $f$ oxides \cite{watson2014occupation}. Therefore, in our collinear magnetism approach, to keep the magnetic ordering as determined at 0~GPa, and at the same time isolate energy changes due only to structural transformations for $P>0$~GPa, we specified for a given Mn ion in all polymorphs the same diagonal unitary occupation of its $d_m$ orbitals ($m=-2,-1,\dots,2\Rightarrow 5\times5$ matrices) using the ``occupation-matrix-control-in-VASP'' algorithm   \cite{watson2014occupation}. In this way, our formerly problematic cases were able to converge. 
\begin{figure*}
\includegraphics[width=0.8\linewidth]{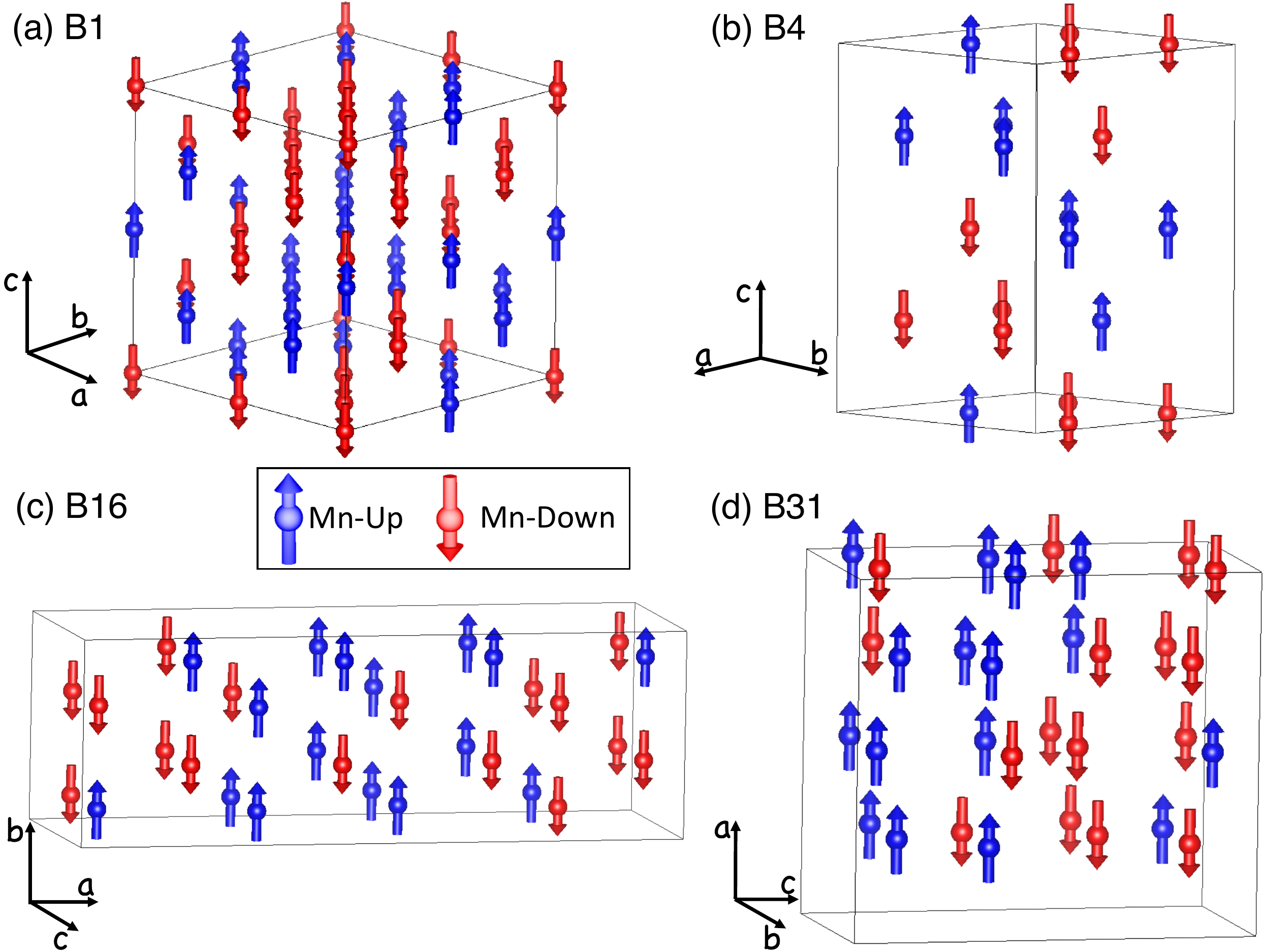}
\caption{\label{pmpolys}Generated SQS supercells representing the PM state of MnS polymorphs as an alloy of randomly distributed Mn $\uparrow$ and $\downarrow$ magnetic moments for (a) cubic B1, (b) hexagonal B4, and orthorhombic (c) B16 and (d) B31 structures. S atoms are not shown for simplification. Structures are visualized using VESTA \cite{vesta}.}
\end{figure*}
\section{\label{}RESULTS AND DISCUSSION}
\subsection{\label{}AFM ordering in MnS }
Firstly, our static DFT+$U$ optimized MnS polymorphs with the AFM ordering considered for each structure are shown in Fig.~\ref{afmpolys}. Our predicted lattice parameter $a_0=5.172$~\AA~for B1-MnS (Fig. \ref{afmpolys}(a)) is only $\sim$0.8\% smaller than its experimental counterpart of 5.212~\AA~ \cite{hastings1956magnetic}. As it was mentioned in Section I, the AFM-II onset in the B1-MnS structure leads to a trigonal distortion, the degree of this distortion can be estimated by the magnitude of the cube corner angle given by $\frac{\pi}{2} + \Delta$, where $\Delta$ measures  the deviation from the ideal cubic symmetry. In our relaxed B1-MnS polymorph we find $\Delta \approx 0.095^{\circ}$, which is in good agreement with the observed deviation of $0.099^{\circ} \pm 0.015^{\circ}$ \cite{morosin1970striction}. For the hexagonal B4-MnS polymorph we obtain the lattice parameters $a_0=3.963$~\AA~and $c_0=6.437$~\AA, which are in excellent agreement with experimental values of $a_0=3.987$~\AA~and $c_0=6.438$~\AA~\cite{hastings1956magnetic}. The optimized volumes per f.u. for each MnS polymorph are listed in Table~\ref{table1} for comparison, and show that B4-MnS is the least dense. 
\begin{figure*}[t!]
\includegraphics[width=0.8\linewidth]{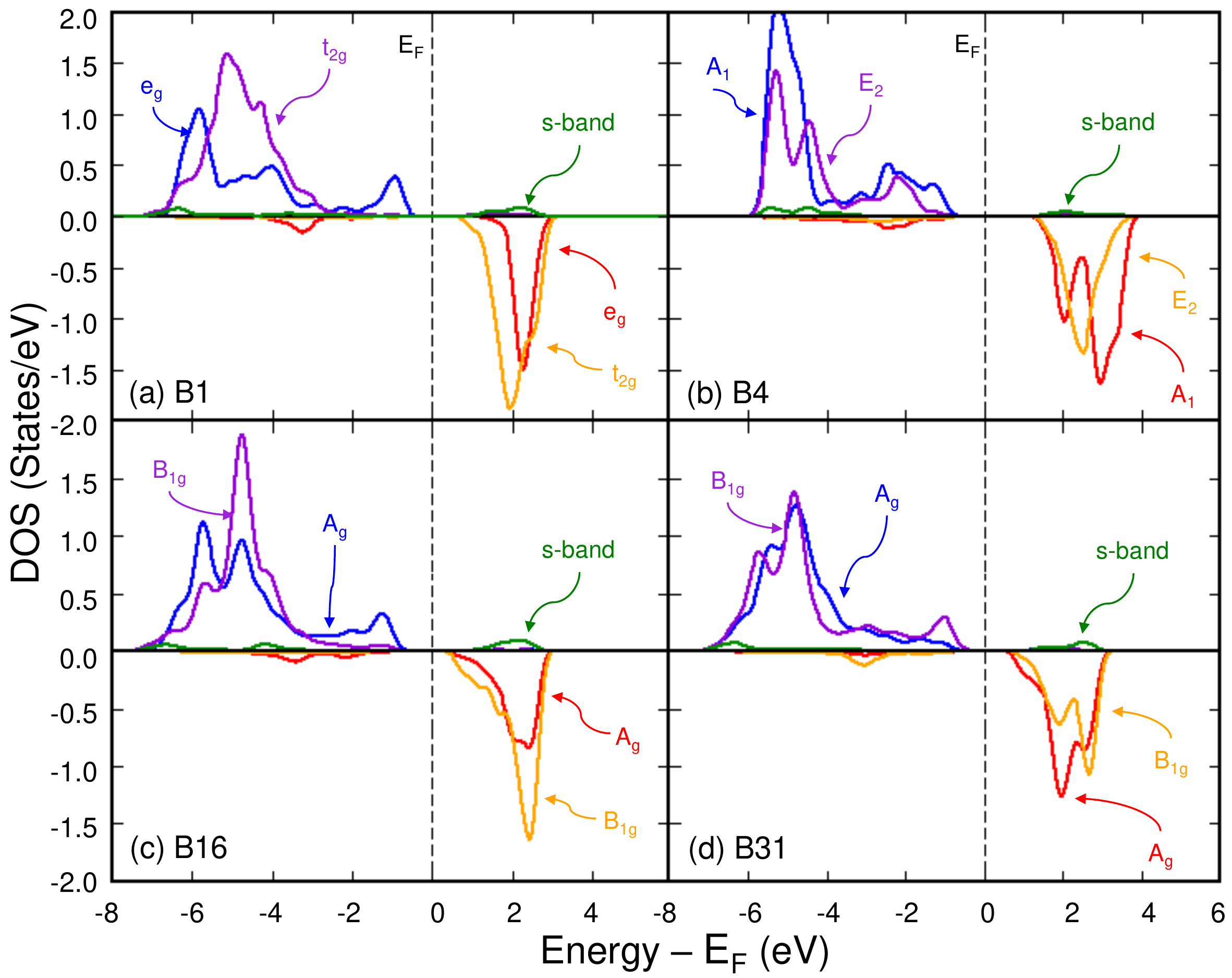}
\caption{\label{pmdos}
DFT$+U$ partial density of states (DOS) for a Mn ion projected on the majority spin (positive) and minority spin (negative) channels with $U=3$~eV within PM SQS's for (a) B1, (b) B4, (c) B16 and (d) B31 MnS polymorphs. The Fermi energy ($E_F$) level is represented by the dashed line at zero.}
\end{figure*}
\begin{table}[h]
\caption{\label{table1}Computed volume $V$ per formula unit (f.u.), energy gap $E_g$, and local magnetic moment $m_{loc}$ of MnS polymorphs with AFM configurations.}
\begin{ruledtabular}
\begin{tabular}{lllccc}
Magnetic & \multicolumn{2}{l}{Polymorph} & $V$/f.u.  & $E_{g}$  & $m_{loc}$   \\
Ordering & & & (\AA$^3$) & (eV) & ($\mu$B) \\

\colrule
AFM-II & RS & B1   & 34.58  & 2.0 & 4.45 \\
AFM-III & WZ & B4  & 43.76 & 2.4 & 4.43  \\
AFM & GeS & B16 & 34.66 & 2.1 & 4.45  \\
AFM & MnP & B31 & 34.38 & 1.6 & 4.42  \\

\end{tabular}
\end{ruledtabular}
\end{table}
%
%
As it is well known, computed DFT band gap energies ($Eg$) are ordinarily underestimated, but from our simulations we find all MnS polymorphs clearly insulating, as can be seen from $E_g$ values in Table~\ref{table1}. For B1-MnS we obtain $E_g=2.0$~eV, which is to some extent, in better agreement with the experimental value of about 3.1~eV \cite{huffman1967optical}, than reports of $E_g=1.36$~eV as predicted from KKR-CPA \cite{oguchi1983band}, $\sim$1~eV from Perdew-Wang (PW) GGA  without $U$ \cite{hobbs1999magnetism}, and $\sim$1.5~eV from GGA-PW$+U$, with $U=3$~eV \cite{rohrbach2003electronic}. Overall, our findings imply that the magnitude of $E_g$ increases with $V/$f.u., with the hexagonal B4-MnS polymorph having the largest energy band gap, while the orthorhombic B31-MnS structure the most narrow (Table~\ref{table1}). 

The projected density of states of Mn $s$ and $d$ orbitals for all MnS polymorphs are shown in Fig.~\ref{afmpdos}. The band characters are assigned roughly by looking at the $m$, $l$ orbital quantum numbers and considering the crystal-field splitting observed for the point group symmetries of the respective structures. For simplicity in this symmetry analysis, we do not consider spin and magnetic anti-unitary operator. Of course,  one could do a complete formal analysis using magnetic point groups and magnetic irreducible co-representations \cite{bradley2009mathematical} if one were after a thorough understanding of the bands, but for the purpose of our study it is not necessary. It should be noted, however, that the onset of the AFM ordering lowers the site symmetry, leading to additional lifting of orbital degeneracies. For example, the $d$-states of octahedrally coordinated Mn$^{2+}$ in B1-MnS split into two levels, a high energy doublet ($e_{g}^{2}$) and a low energy triplet ($t_{2g}^{3}$). Here, the band gap is opened between $e_g$--majority and $t_{2g}$--minority bands, Fig. 2(a). However, the onset of AFM-II ordering lowers the symmetry from octahedral $O_h$ to rhombohedral $D_{3d}$, resulting in additional splitting of the $t_{2g}$ states into a singlet and a doublet. In B4-MnS the hexagonal crystal field splits the $d$-shell into a singlet $A_{1}$ and two doublets $E_1$ and $E_2$. According to our calculation, in this structure the band gap is opened between the $A_{1}$--majority singlet and the $E_2$--minority doublet, Fig. 2(b), while the AFM-III ordering lowers the original $C_{6v}$ site symmetry to $C_{1h}$. Lastly, both B16- and B31-MnS structures belong to the same $D_{2h}$ point group. The orthorhombic crystal field of this symmetry lifts the five-fold degeneracy of the spherical $d$-shell leading to five singlets. Fig.~\ref{afmpdos}(c) and (d) show that the band gap is opened between $B_{1g}$--majority and $A_g$--minority bands in both phases. \\
Our calculated local magnetic moments of the four AFM-MnS polymorphs are very similar in magnitude as one can see from Table~\ref{table1}. For RS-MnS we find a $m_{loc}=4.45$~$\mu$B, which is in much better correspondence with experimental observations of 4.54~$\mu$B  \cite{fender1968covalency} than earlier calculations of 4.92 ~$\mu$B under a Hartree-Fock scheme \cite{tappero1997electronic}, 4.39~$\mu$B  from KKR-CPA \cite{oguchi1983band}, and 4.082 ~$\mu$B within GGA-PW \cite{hobbs1999magnetism}. 
\subsection{Magnetically disordered PM MnS}
Our converged MnS polymorphs in PM states as modeled by the construction of SQS's are shown in Fig.~\ref{pmpolys}. Here we emphasize the fact that in our calculations, all degrees of freedom (lattice parameters and ionic positions) were allowed to relax, unlike other studies using SQS structures for TM oxides in which the symmetry and volume were kept fixed \cite{zunger2018polymorphous}. At first sight, these constrictions may appear reasonable to apply, as experimentally, there is no observable distortion from the cubic symmetry of the PM B1-MnS polymorph. 
Yet, allowing full relaxations in ionic coordinates and lattice parameters of our SQS structures and comparing to their respective ideal symmetry could be a criterion to judge how {\it well} the created SQS represents its polymorph. For example, for PM B1-MnS (Fig.~\ref{pmpolys}(a)) we find, as in the case when AFM ordering is achieved, a small trigonal distortion. The computed deviation from the ideal cubic angle for our 64-atom SQS supercell is $\Delta \approx 0.024^{\circ}$, whereas for a 216-atom SQS supercell we obtain $\Delta \approx 0.003^{\circ}$, one order of magnitude smaller than for the 64-atom SQS supercell. Using the B1 64-atom and 216-atom SQS's we find lattice parameters of $a_0=5.181$~\AA~and 5.1807~\AA~, respectively. These values are somewhat underestimated, but still in good agreement, with respect to measurements of 5.225~\AA~\cite{mccammon1991static}, 5.225~\AA~\cite{sweeney1993compression}, and 5.29~\AA~\cite{b31}.
Similarly to the AFM-ordered MnS phases, we find all PM MnS polymorphs to be insulators, although with smaller energy band gaps (Table~\ref{table2}) than their AFM counterparts (Table~\ref{table1}). . Our PM  $E_g$ values (Table~\ref{table2}) suggest, however, that magnetic disorder affects the least the energy band gap of the hexagonal B4-MnS polymorph in comparison to its AFM analog, while B16-MnS shows the largest $E_g$ reduction. We particularly highlight the fact that, even though for B1-MnS our predicted $E_g=1.4$~eV is fairly underestimated as compared to the observed values of  2.7~eV \cite{sato1997chemical} and 2.8~eV \cite{huffman1967optical}, we obtain an insulating PM B1-MnS phase, unlike other calculations in which it was found to be metallic \cite{oguchi1983band,raybaud1997ab}.
\\
Finally, examining our calculated local magnetic moments in the PM phases listed in Table~\ref{table2}, we observe a slight increase in magnitude when compared to the results in the AFM structures (Table~\ref{table1}).  For the PM B1-MnS  our predicted value is 4.50~$\mu$B.  

\begin{table}[h]
\caption{\label{table2}Computed volume $V$ per formula unit (f.u.), energy gap $E_g$, and local magnetic moment $m_{loc}$ of MnS polymorphs in PM state.}
\begin{ruledtabular}
\begin{tabular}{lllccc}
Magnetic & \multicolumn{2}{l}{Polymorph} & $V$/f.u.  & $E_{g}$  & $m_{loc}$   \\
Ordering & & & (\AA$^3$) & (eV) & ($\mu$B) \\
\colrule
PM & RS & B1   & 34.76  & 1.4 & 4.50 \\
 & WZ & B4    & 44.06 & 2.3 & 4.44  \\
 & GeS & B16  & 36.94 & 1.2 & 4.50  \\
 & MnP & B31  & 34.94 & 0.9 & 4.46 \\
\end{tabular}
\end{ruledtabular}
\end{table}
%
To understand better the decrease in the magnitude of the PM energy band gaps with respect to the AFM cases, we show in Fig.~\ref{pmdos} the projected density of states of a Mn ion in the PM state of the four MnS  polymorphs. Our calculations show that the introduction of magnetic disorder in the MnS polymorphs {\it spreads and produces} new spin-majority and spin-minority states in the $E_g$ range of the AFM ordered phases (Fig.~\ref{afmpdos}). Consequently, magnetic disorder exclusively (as the occupation matrix is constrained) {\it forces} a shift of the Fermi energy level in reference to the ordered AFM cases, but in such a way that the resulting PM MnS phases remain insulating and hence with smaller energy band gaps. In principle, different magnetic disorder given by distinct SQS's would give rise to different energy band gap openings, this effect was reported in the context of NbMnSb by disordering Mn and Ni with respect to the sites they occupy in the ordered phase \cite{orgassa1999first}. Additionally, in Fig.~\ref{pmdos}(a) and (c) we can see that for PM B1- and B31-MnS phases, the energy band gap opens, as in their AFM analogs, between the $e_g$--majority and the $t_{2g}$--minority bands, and between $B_{1g}$--majority and $A_g$--minority, respectively. However, for PM B4- and B16-MnS phases, magnetic disorder changes the character of the band gap openings in reference to their AFM counterparts, to be between $A_{1}$--majority and $A_1$--minority (Fig.~\ref{pmdos}(b)), and $A_{g}$--majority and $B_1$--minority  (Fig.~\ref{pmdos}(d)), respectively.
\subsection{AFM to PM evolution of the band gap}
\begin{figure}[t!]
\includegraphics[width=0.9\linewidth]{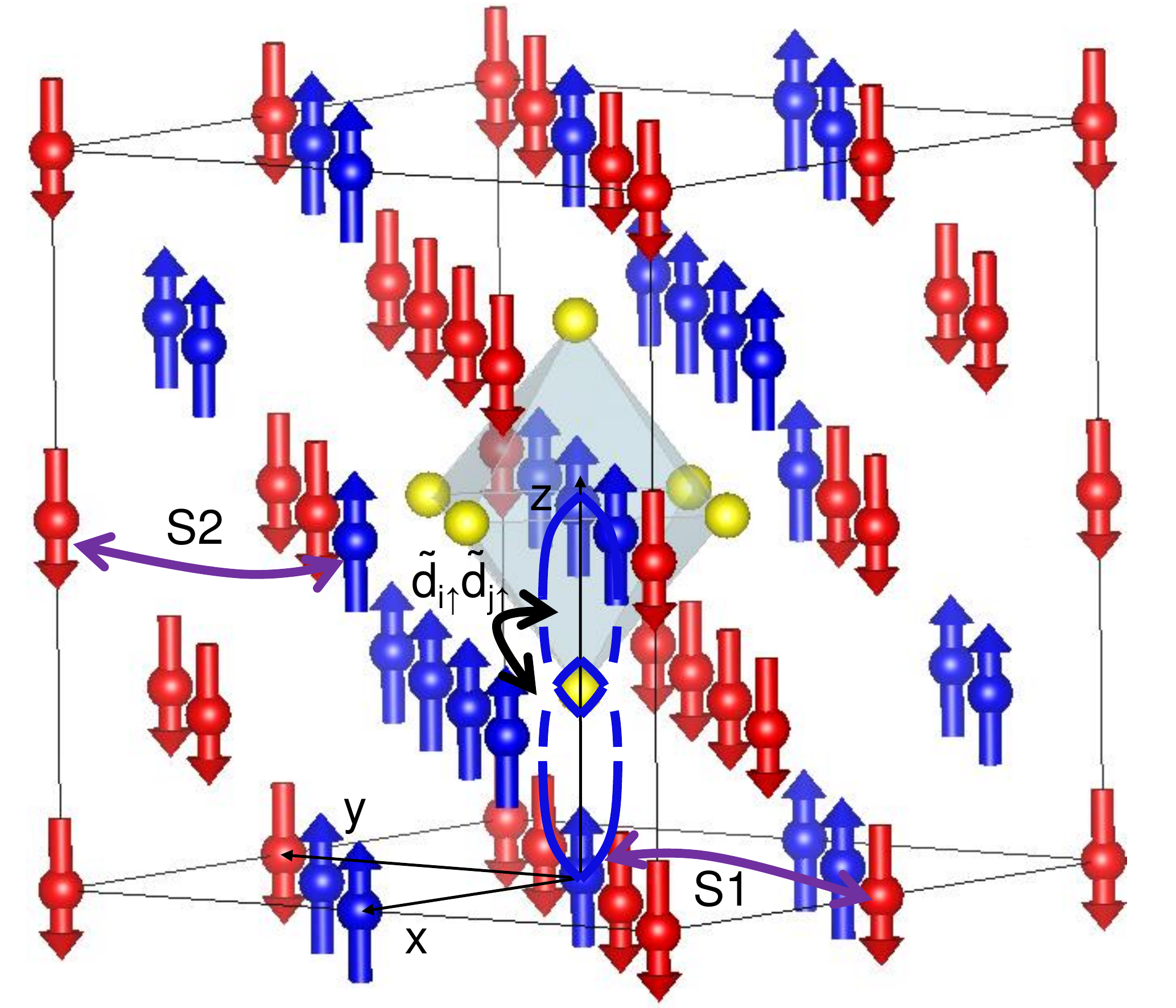}
\caption{Representation of {\it slight} deviations from the AFM-II ordering. Blue-up and red-down arrows are Mn atoms with  $\uparrow$- or $\downarrow$-magnetic moment, yellow balls are S atoms. The violet lines indicate two cases of introducing ``magnetic disorder" by swapping Mn atoms of opposite magnetic moments. In the text, a structure  with D1-disorder  corresponds to having only one pair of Mn-$\uparrow$ and Mn-$\downarrow$ swapped (S1), whereas a structure with  D2-disorder  has swaps S1 and S2 as indicated. Half-solid-half-dashed loops depict $p$-$d$ hybridized orbitals. When two Mn atoms with parallel magnetic moments are coupled through a 180$^o$ Mn$\uparrow$-S-Mn$\uparrow$ bond, an additional coupling between $d$-derived orbitals also occurs $\tilde{d}_{i\uparrow}^{\dagger}\tilde{d}_{j\uparrow}$.}\label{D1D2}
\end{figure}
\begin{figure*}[t!]
\includegraphics[width=0.8\linewidth]{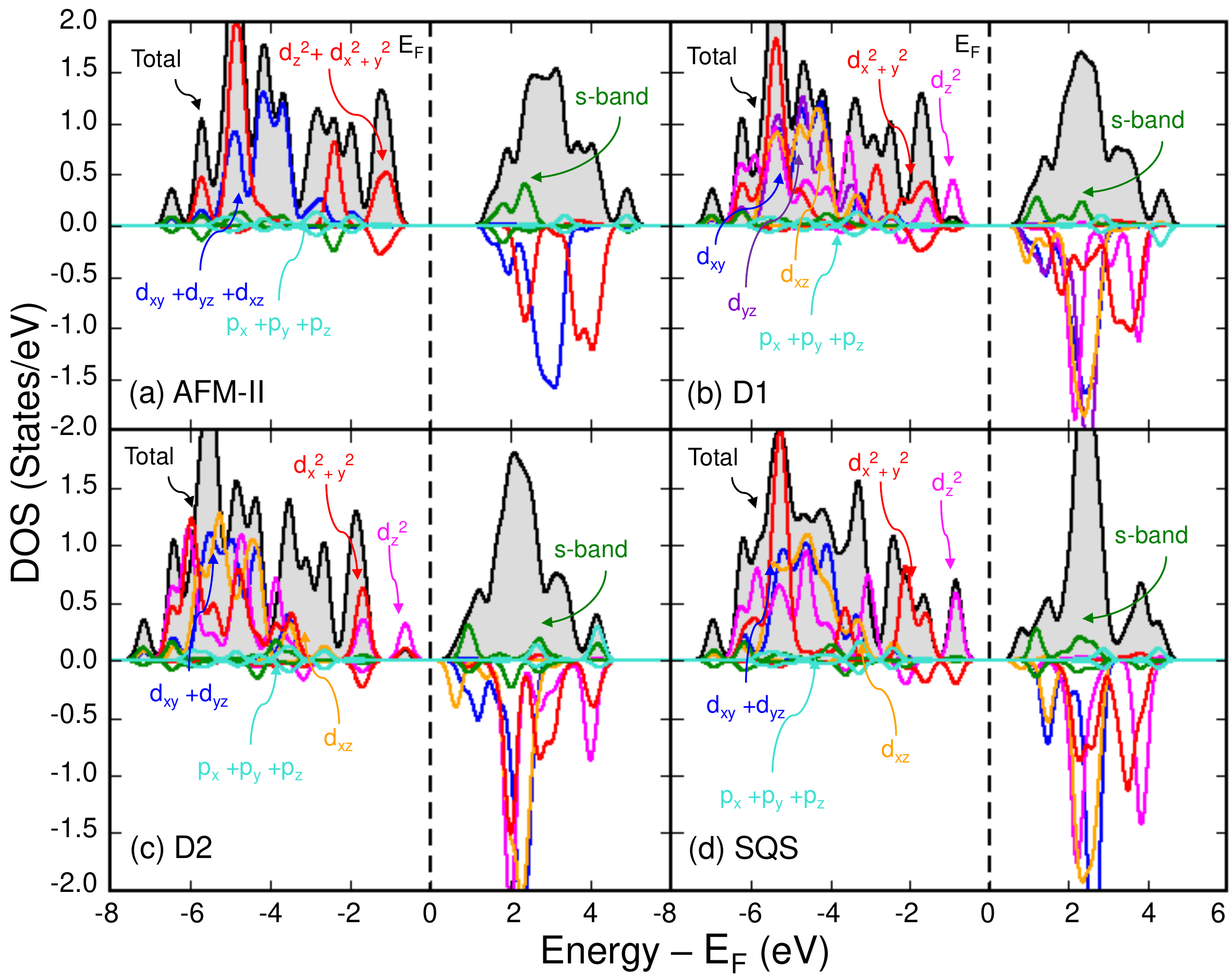}
\caption{Orbital-projected density of states on the majority spin (positive) and minority spin (negative) channels with $U=3$~eV of four different configurations of the B1-MnS with increasing level of disorder. The magnitude of the $p$-states is magnified by a factor of three for visibility. (a) The DOS of the AFM state is largely determined by the $p$-$d$ and $s$-$d$ hybridizations between S and Mn centered orbitals. (b) and (c) The shrinkage of the band gap in the spin-majority channel is mainly caused by the formation of a 180$^o$ Mn-$\uparrow$-S-Mn-$\uparrow$ chain that is absent in the AFM case. Additionally, there is also splitting in the $t_{2g}$-derived states. This is mostly determined by the overall number of Mn atoms with parallel magnetic moment in a given plane. For example, the structure D2 has the same number of Mn-$\uparrow$ atoms in $xy$ and $yz$ planes which is why we observe degeneracy in these states, whilst this number is different in all planes in D1. (d) The fully random SQS structure exhibits the features of D1 in the $e_g$ splitting whilst retaining the $xy+yz$ degeneracy shown in D2.}\label{disorderDOS}
\end{figure*}
In order to gain more insights into the narrowing of the band gap of MnS in going from an AFM ordered phase to its magnetically disordered PM one above its $T_N$, we {\it gradually introduce disorder} in steps. We accomplish this task by studying in more detail the projected Mn DOS for a couple of configurations slightly departing from the original AFM ordering.
First, we recall that the most general non-interacting Hamiltonian within the DFT+$U$ scheme is given by \cite{hubbard1963electron,bruus2004many}:
\begin{equation}
\begin{split}
    \hat{H} = \sum_{i,j,m,m^{\prime},\sigma}\big(t_{i,j,\sigma}^{m,m^{\prime}}c^{\dagger}_{im,\sigma}c_{jm^{\prime},\sigma} + h.c.\big) \\
+ U\sum_{i,m^{\prime\prime}}n_{im^{\prime\prime}\sigma}<n_{im^{\prime\prime}-\sigma}>, \label{NIHam}
\end{split}
\end{equation}
where $c^{\dagger}_{im,\sigma}$ and $c_{jm^{\prime},\sigma}$ create and destroy an electron with spin $\sigma$ on site $i$, orbital $m$, and site $j$, orbital $m^{\prime}$, respectively. $t_{i,j,\sigma}^{m,m^{\prime}}$ are the hybridization integrals, which one can assume to follow the two-center Slater-Koster (SK) approximation \cite{slater1954simplified} as $t_{i,j,\sigma}^{m,m^{\prime}} \approx E_{i,j,\sigma}^{m,m^{\prime}}$, i.e., hybridization between orbital $m$ of character $p_i$ and orbital $m^{\prime}$ of character $d_j$.
The last term in Eq. (\ref{NIHam}) represents the mean-field decoupling of the Hubbard interaction term and only acts on the $d$-states of the same site ($m^{\prime\prime}$ runs only through $d$-orbitals), with $n_{im^{\prime\prime}\sigma}$ being the number operator ($c^{\dagger}_ic_i$) and $<n_{im^{\prime\prime}-\sigma}>$ the average occupation of the respective orbital and spin (in our work, these quantities are fixed in the DFT cycle using the OMC).

\begin{figure}[t!]
\includegraphics[width=0.7\linewidth]{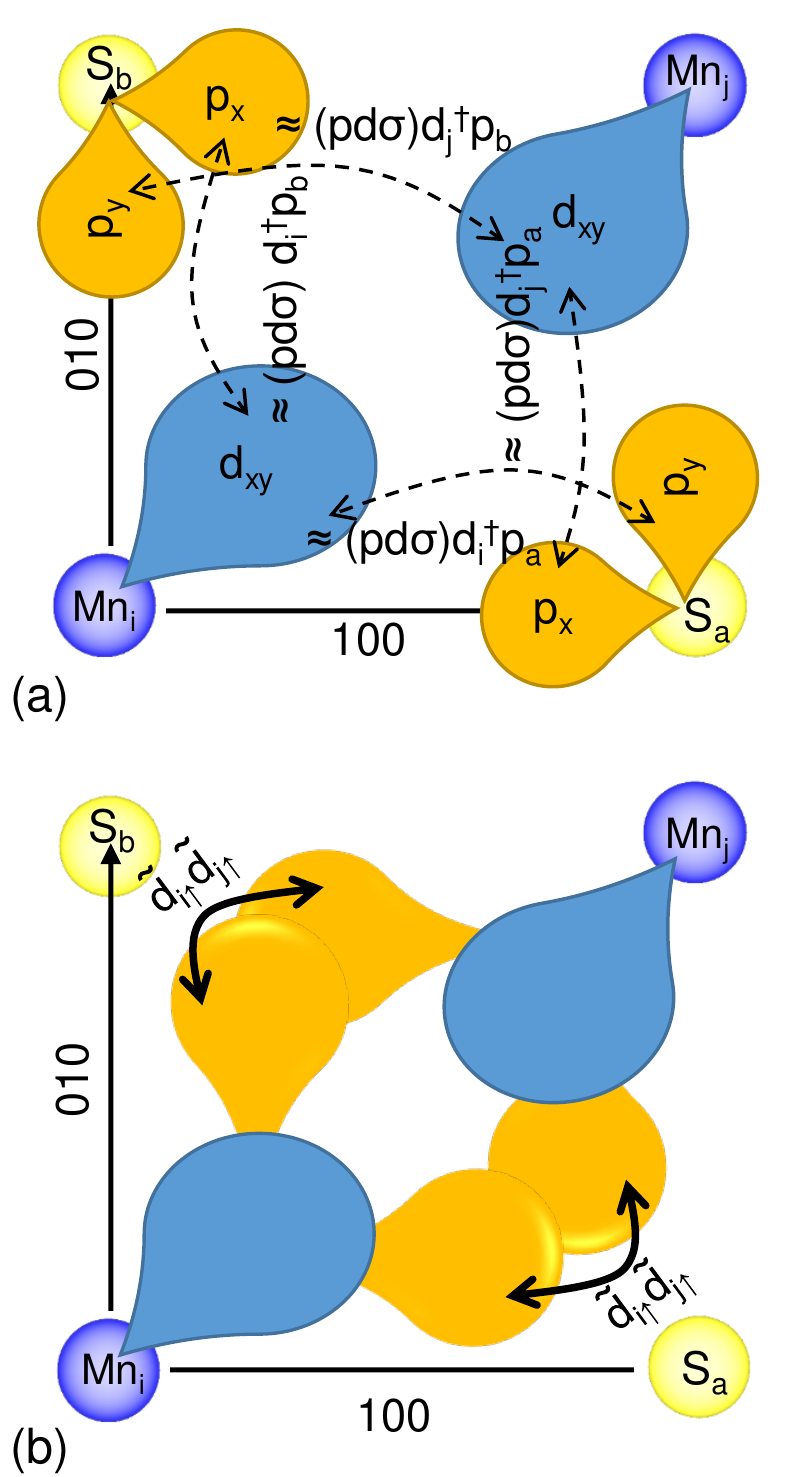}
\caption{Schematic coupling between $d$-derived orbitals of neighboring Mn sites. (a) In AFM-II B1-MnS, $d$-orbitals hybridize with $p$-orbitals of sulfur ions S$_a$ and S$_b$ in the (010) and (100) directions, respectively. (b) Nearest Mn neighbors with parallel magnetic moments allow for an additional coupling between $p$-$d$ hybridized orbitals ($\tilde d_{i\uparrow}\tilde d_{j\uparrow}$). When the number of Mn atoms with parallel magnetic moment in the $xy$-, $yz$- or $xz$-plane is changed through disorder, the respective orbital degeneracy is lifted.\label{ddhyb}
}
\end{figure}

Below we present our results focused solely on the B1-MnS structure, although for the other three polymorphs a similar analysis would also apply. 
Thus, we {\it slightly disorder} the AFM-II 64-atom supercell of B1-MnS and investigate two cases before achieving a fully disordered PM structure. Firstly, we consider a structure in which in a 180$^o$ chain, we swap a couple of Mn atoms with antiparallel magnetic moments (D1-disorder), and then a structure with two of such swaps (D2-disorder), as shown in Fig. \ref{D1D2}. For simplification, we exclude long range interaction effects between neighboring unit-cells. To isolate the system's reaction to disorder, we also suppress all relaxation effects and we only use the $\Gamma$-point in the self-consistent calculation cycle.
Taking the converged ground state for the AFM-II B1-MnS phase as the reference state, in which the largest non-zero hybridization integrals in Eq. \ref{NIHam} arise between Mn $d$-orbitals, and $s$- and $p$-orbitals of S (Fig. \ref{disorderDOS}(a)), we observe that upon introducing D1-disorder (i.e., after the creation of a FM coupling in one of the 180$^o$ chains, Fig. \ref{D1D2}), there is now a possibility of additional combination between $p$-$d$ hybridized orbitals of neighboring Mn sites with parallel magnetic moments. This {\it re-hybridization} seems to greatly contribute to the lift of the degeneracy of the $t_{2g}$- and $e_g$-derived states and the spreading and pushing of the latter at the top of the valence band, effectively  shrinking the energy gap as shown in Fig. \ref{disorderDOS}(b). The effect of this re-hybridization in the {\it shift} of $e_g$ states is more pronounced when D2-disorder (with S1 and S2 swaps) is considered  (Fig. \ref{disorderDOS}(c)). However, once that a number of swaps occurs randomly and the PM state is reached in the SQS of B1-MnS, the energy band gap reaches its final magnitude. Another possible mechanism of re-hybridization, also partially responsible for the lift of the $t_{2g}$ degeneracy, can take place within planes. In the AFM-II Mn $d$-orbitals hybridize with $p$-orbitals of S ions located at 90$^o$. That is, according to SK rules, in octahedral geometry $d_{xy}$ orbitals hybridize only with $p_x$ and $p_y$ in the (010) and (100) directions, as shown in Fig.~\ref{ddhyb}(a). However, when the nearest Mn neighbors have parallel magnetic moments, there is an additional coupling between the $p$-$d$ hybridized orbitals as depicted in Fig.~\ref{ddhyb}(b). Such re-hybridization is made possible through the admixture of $p$-states that provide the necessary {\it spatial extension} of the otherwise highly localized $d$-states. These re-hybridized $\tilde{d}$-orbitals centered on the Mn atoms can further mix with each other provided there is enough spatial overlap between them. In such a case, the $t_{2g}$ degeneracy is predominantly determined by the number of Mn atoms with parallel magnetic moments in $xy$, $yz$ and $xz$ planes, i.e., if this number is different for two given planes the degeneracy between the respective orbitals is lifted, Fig.~\ref{disorderDOS}(b)-(c).

\subsection{\label{}High-pressure landscape of MnS polymorphs}
\begin{figure}
\includegraphics[width=1\linewidth]{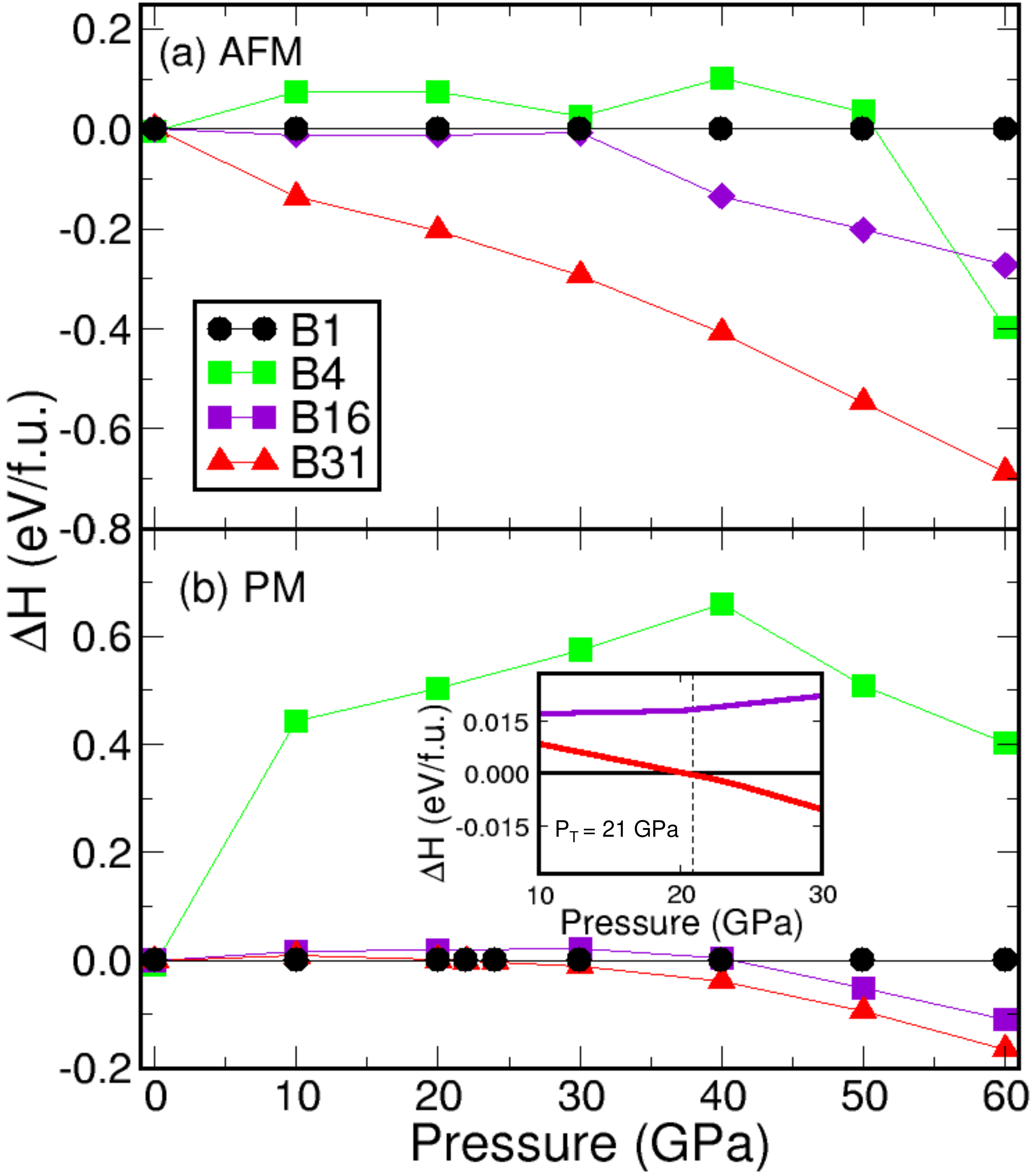}
\caption{\label{figH}Static PBE+$U$  relative enthalpy $\Delta H$ per f.u. between different  (a) AFM and (b) PM states of MnS polymorphs. All results are given in reference to the B1 structure.}
\end{figure}

As discussed in the Introduction, room temperature experimental studies of the stability of the B1-MnS polymorph under pressure, have reported a structural change to a new phase. However, the results were not conclusive as the transition pressure and the new phase were not uniquely determined  \cite{kraft1988high,mccammon1991static,sweeney1993compression}. 
Therefore, in order to examine the likelihood of a pressure induced structural phase transformation of B1-MnS, for each MnS polymorph we computed the enthalpy ($H$) as a function of pressure ($P$), volume ($V$), and internal energy ($E$), i.e., 
\begin{equation}
H(P)=E[V(P)]+PV(P). \label{eqH}
\end{equation}
In Fig.~\ref{figH} we show the relative enthalpy $\Delta H$ per f.u. of the MnS polymorphs studied here with respect to the B1-MnS structure, i.e., $\Delta H=H(\text{MnS-phase})-H(\text{B1})$, as a function of pressure between 0 and 60~GPa. Although high-pressure experiments at $T_R$ deal with the PM phases of MnS, we present results for both AFM and PM polymorphs.
\begin{figure}[ht]
\includegraphics[width=1\linewidth]{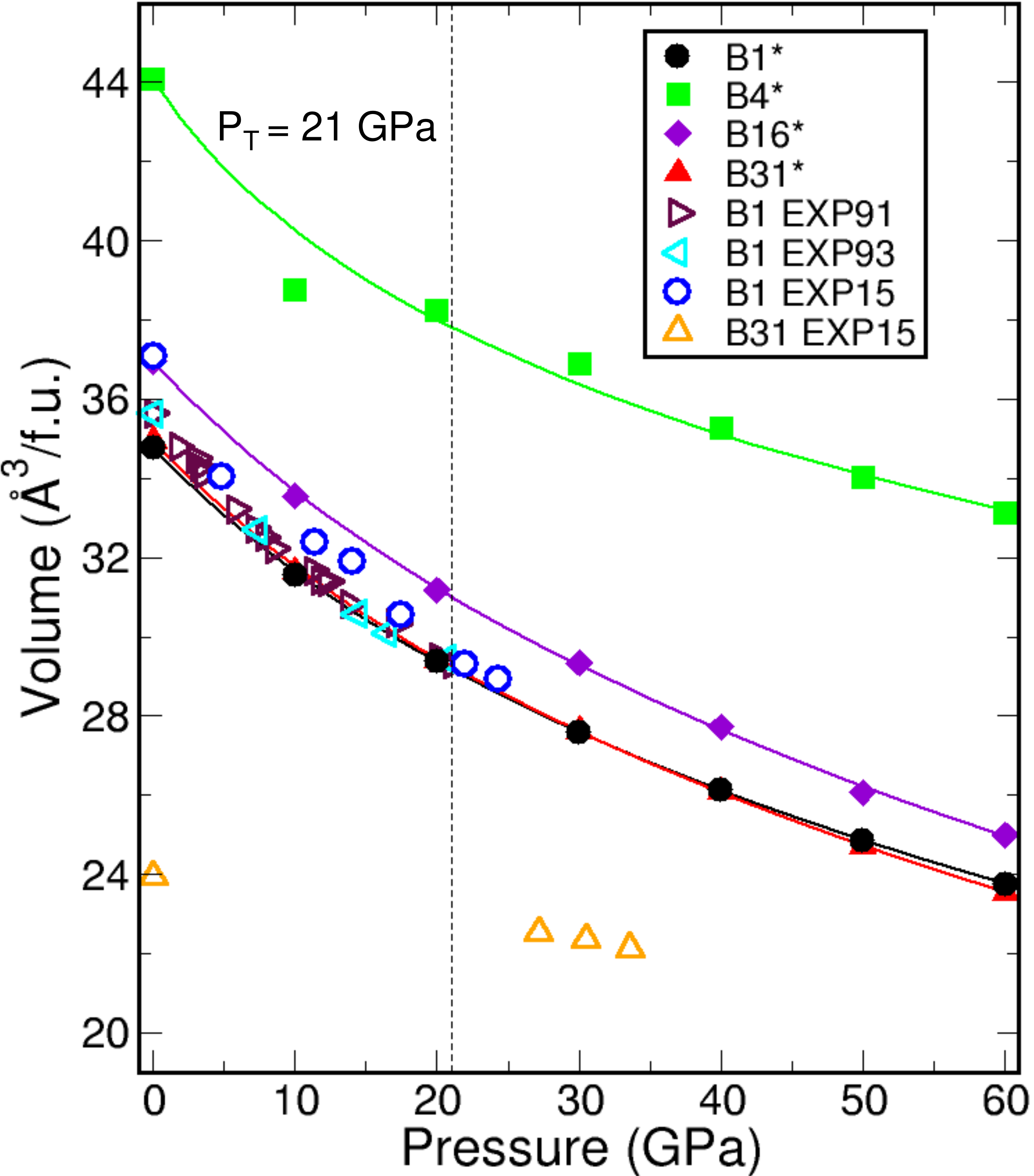}
\caption{\label{volpres}Pressure dependence of volume per f.u. for all PM MnS polymorphs as predicted by our computations. We compare our PM MnS results [$\star$] to experimental values: EXP91 \cite{mccammon1991static}, EXP93 \cite{sweeney1993compression}, and EXP15 \cite{b31}. The dashed line indicates our predicted transition pressure ($P_T=21$~GPa) at which B1$\rightarrow$B31. Solid lines are fits of the 3rd order Birch-Murnaghan's equation of state to our calculated volumes.}
\end{figure}
On one hand, our static $\Delta H$ values suggest that if we were at a sufficiently low temperature to achieve AFM ordering for all MnS polymorphs, B31-MnS would be the most stable phase at 0~GPa, and it would remain so up to 60~GPa, Fig.~\ref{figH}(a). On the other hand, from our static calculations of the enthalpy in PM phases of MnS, Fig.~\ref{figH}(b), we conclude that the RS structure B1-MnS is the most stable phase at ambient conditions, but as pressure increases, it undergoes a transformation to the orthorhombic B31-MnS polymorph at $P_T\approx21$~GPa, as illustrated for clarity by the inset in Fig.~\ref{figH}(b). Our predicted $P_T$ in bulk B1-MnS  is in close agreement with the experimentally observed structural transformation of B1-MnS to an unidentified phase with lower symmetry than hexagonal (B4-MnS) at $\sim$26~GPa \cite{sweeney1993compression}. Our determined B1$\rightarrow$B31 transformation was also established at about 22.3~GPa in experiments synthesizing high-pressure MnS nanorods \cite{b31}, in which these B31-MnS nanorods were quenchable to 0~GPa. Furthermore,  static LDA+$U$ DFT calculations (with an effective $U=5.13$~eV) performed alongside the experimental study found that, although the B1- and B31-MnS polymorphs are energetically very close below 8~GPa, B31 is the most stable at all pressures between 0 and 40~GPa. However, unlike in our static PBEsol calculations using the constructed magnetically disorder SQS PM states of MnS, the former LDA+$U$ modeling \cite{b31} does not specify how the PM was simulated.
We notice as well, that our four PM polymorphs considered here, at the level of our calculations, are remarkably close in energy at 0~GPa, but the B1 possesses the absolute lowest energy per f.u., and as pressure increases, B16 and B31 enthalpies increase slightly and then decrease to start competing for stability against the B1-MnS structure, with B31-MnS eventually becoming more stable at $P_T\approx21$~GPa. We also do not observe any other further transformation up to 60~GPa.  The predicted behavior from our computations for the AFM MnS polymorphs' enthalpy trends are surprisingly in qualitative agreement  with the LDA+$U$ modeling of the B1- and B31-MnS phases \cite{b31}. 
%
\begin{figure}[ht]
\includegraphics[width=1\linewidth]{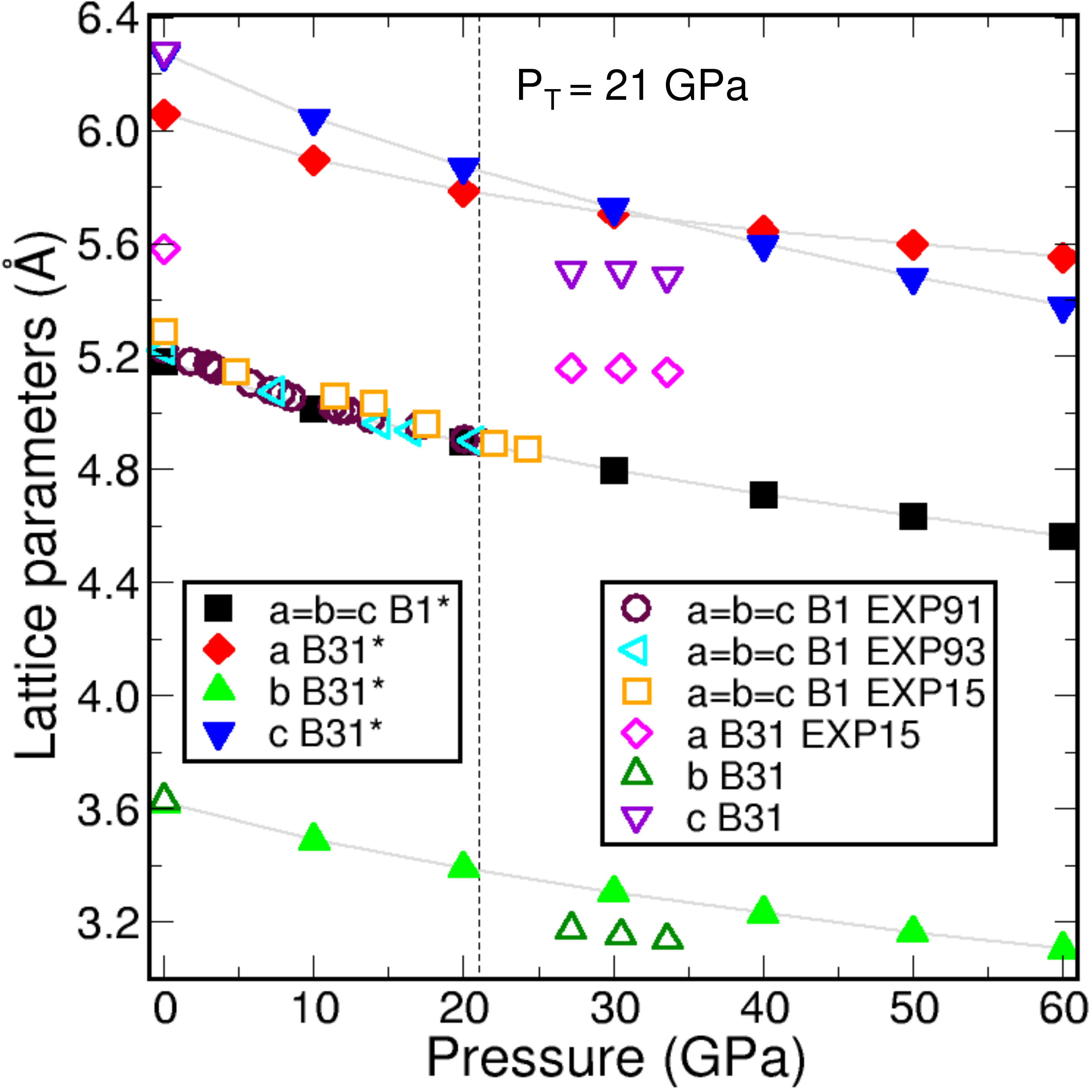}
\caption{\label{lattpar}Lattice parameters as a function of pressure for cubic B1- and orthorhombic B31-MnS structures. Static PBE+$U$ results [$\star$] of our magnetically disordered PM polymorphs are compared to experimental values: EXP91 \cite{mccammon1991static}, EXP93 \cite{sweeney1993compression}, and EXP15 \cite{b31}. The dashed line indicates our predicted transition pressure ($P_T=21$~GPa) at which B1$\rightarrow$B31.}
\end{figure}
%
In Fig.~\ref{volpres} we show our predicted change in volume per f.u. as pressure increases for the PM MnS structures modeled in this study. We can see that experimental volumes for cubic B1-MnS \cite{mccammon1991static,sweeney1993compression} are exceptionally consistent with our findings. The largest deviation from our computed $V/$f.u. occurs for B31-MnS with respect to measurements provided from high-pressure experiments on MnS nanorods \cite{b31}. As a consequence of the B1$\rightarrow$B31 structural transformation, from our calculated trends we obtain a decrease in volume of MnS of only approximately 2\% at $P_T$. This volume reduction, however, is one order of magnitude smaller in contrast to the one observed in the B31 nanorods \cite{b31}. 

A closer look at the pressure dependance of our predicted lattice parameters for the B1-MnS shows that our calculated values are  in excellent agreement with available experimental data \cite{mccammon1991static,sweeney1993compression,b31}, as can be seen in Fig.~\ref{lattpar}. In the case of B31-MnS, our calculated $a,b,$ and $c$ lattice constants are systematically overestimated by about $\sim$10\%, $\sim$5\% and $\sim$4\%, respectively, in comparison to experimental values at high pressures \cite{b31}. At 0~GPa, the agreement between our results and the experimental values becomes much better for $a_0$ and $b_0$, but curiously our $c_0$ magnitude is $\sim$8\% larger than the only measurement reported up to date \cite{b31}. 

Lastly, we obtain the bulk modulus at 0~GPa ($K_0$) from Birch-Murnaghan's third order equation of state (EOS) fittings to our $P-V$ data between 0 and 60~GPa. Our predicted $K_0\approx92$~GPa (with its pressure derivative $K'_0=3$) for PM B1-MnS is in reasonable agreement with the experimental value reported of 88$\pm$6~GPa as fit with a variable $K'_0$ \cite{sweeney1993compression}. Our result is, however, overestimated by almost 20\%, when the experimental data is fit using a constant $K'_0=4$. A comparison between our computed bulk moduli indicates that B1-, B16-, and B31-MnS polymorphs oppose almost indistinguishable resistance against volume compression under hydrostatic pressure, while B4 is the easiest to compress with a $K_0\approx57$~GPa.

\section{CONCLUSIONS}
In summary, we carried out static first-principles calculations to model the AFM and PM states of four MnS polymorphs. We demonstrated that the combination of PBE+$U$ with the construction of SQS supercells and localization of the Mn $d$-electrons through occupation matrix control methods allowed us to achieve not only convergence and accuracy of structural optimizations but was vitally crucial to obtain finite energy band gaps and local magnetic moments in the PM phases. This result was particularly important in the case of the PM rock-salt B1-MnS polymorph, which experimentally has been observed to be an insulator, but was predicted to be metallic by multiple simulations. In our study we also showed that with our approach, we were able to isolate energy changes as a function of pressure due purely to ionic and lattice parameters relaxation under hydrostatic compression. In this manner, we computed the enthalpies of the PM MnS polymorphs in order to explore their high-pressure landscape to detect structural transformations, and indeed, we  determined the cubic B1-MnS structure to be  the most stable at ambient pressure and up to approximately 21~GPa, pressure at which B1 undergoes a structural transformation to the orthorhombic B31-MnS phase. The overall trends in the electronic structure as well as phase transitions are expected to hold for different $U$ values. The increase (decrease) of the $U$ value will result in the increase (decrease) of the structural parameters and, therefore, in the change of magnitude of  the transition pressure. The chosen value in our work was supported by the close agreement to experimental structural parameters. Our predicted B1$\rightarrow$B31 transformation, in the context of all our modeling considerations, is rather meaningful as it closely resembles a structural phase transition observed from  X-ray diffraction and high-pressure experiments performed on B1-MnS, in which the new but unidentified phase was reported at about 26~GPa.  Overall, our approach proved to be accurate in the modeling of manganese sulphide polymorphs, thus we plan to extend it in the future to the investigation of other TM compounds.
\begin{acknowledgments}
The authors gratefully acknowledge the Gauss Centre for Supercomputing e.V. (www.gauss-centre.eu) for funding this project by providing computing time through the John von Neumann Institute for Computing (NIC) on the GCS Supercomputer JUWELS at J\"ulich Supercomputing Centre (JSC) under project {\bf abinitiomodmatsgeo}. AC and MNV were supported by the Helmholtz Association through {\it funding of first-time professorial appointments of excellent women scientists (W2/W3)}
\end{acknowledgments}
\section*{author declarations}
The authors have no conflicts to disclose.

\bibliographystyle{apsrev4-1}
\bibliography{bibliography}

\begin{thebibliography}{44}%
\makeatletter
\providecommand \@ifxundefined [1]{%
 \@ifx{#1\undefined}
}%
\providecommand \@ifnum [1]{%
 \ifnum #1\expandafter \@firstoftwo
 \else \expandafter \@secondoftwo
 \fi
}%
\providecommand \@ifx [1]{%
 \ifx #1\expandafter \@firstoftwo
 \else \expandafter \@secondoftwo
 \fi
}%
\providecommand \natexlab [1]{#1}%
\providecommand \enquote  [1]{``#1''}%
\providecommand \bibnamefont  [1]{#1}%
\providecommand \bibfnamefont [1]{#1}%
\providecommand \citenamefont [1]{#1}%
\providecommand \href@noop [0]{\@secondoftwo}%
\providecommand \href [0]{\begingroup \@sanitize@url \@href}%
\providecommand \@href[1]{\@@startlink{#1}\@@href}%
\providecommand \@@href[1]{\endgroup#1\@@endlink}%
\providecommand \@sanitize@url [0]{\catcode `\\12\catcode `\$12\catcode
  `\&12\catcode `\#12\catcode `\^12\catcode `\_12\catcode `\%12\relax}%
\providecommand \@@startlink[1]{}%
\providecommand \@@endlink[0]{}%
\providecommand \url  [0]{\begingroup\@sanitize@url \@url }%
\providecommand \@url [1]{\endgroup\@href {#1}{\urlprefix }}%
\providecommand \urlprefix  [0]{URL }%
\providecommand \Eprint [0]{\href }%
\providecommand \doibase [0]{http://dx.doi.org/}%
\providecommand \selectlanguage [0]{\@gobble}%
\providecommand \bibinfo  [0]{\@secondoftwo}%
\providecommand \bibfield  [0]{\@secondoftwo}%
\providecommand \translation [1]{[#1]}%
\providecommand \BibitemOpen [0]{}%
\providecommand \bibitemStop [0]{}%
\providecommand \bibitemNoStop [0]{.\EOS\space}%
\providecommand \EOS [0]{\spacefactor3000\relax}%
\providecommand \BibitemShut  [1]{\csname bibitem#1\endcsname}%
\let\auto@bib@innerbib\@empty
\bibitem [{\citenamefont {Tappero}\ \emph {et~al.}(1997)\citenamefont
  {Tappero}, \citenamefont {D'Arco},\ and\ \citenamefont
  {Lichanot}}]{tappero1997electronic}%
  \BibitemOpen
  \bibfield  {author} {\bibinfo {author} {\bibfnamefont {R.}~\bibnamefont
  {Tappero}}, \bibinfo {author} {\bibfnamefont {P.}~\bibnamefont {D'Arco}}, \
  and\ \bibinfo {author} {\bibfnamefont {A.}~\bibnamefont {Lichanot}},\
  }\href@noop {} {\bibfield  {journal} {\bibinfo  {journal} {Chem. Phys.
  Lett.}\ }\textbf {\bibinfo {volume} {273}},\ \bibinfo {pages} {83} (\bibinfo
  {year} {1997})}\BibitemShut {NoStop}%
\bibitem [{\citenamefont {He}\ \emph {et~al.}(2014)\citenamefont {He},
  \citenamefont {Hart}, \citenamefont {Liang}, \citenamefont {Garsuch},\ and\
  \citenamefont {Nazar}}]{he2014stable}%
  \BibitemOpen
  \bibfield  {author} {\bibinfo {author} {\bibfnamefont {G.}~\bibnamefont
  {He}}, \bibinfo {author} {\bibfnamefont {C.~J.}\ \bibnamefont {Hart}},
  \bibinfo {author} {\bibfnamefont {X.}~\bibnamefont {Liang}}, \bibinfo
  {author} {\bibfnamefont {A.}~\bibnamefont {Garsuch}}, \ and\ \bibinfo
  {author} {\bibfnamefont {L.~F.}\ \bibnamefont {Nazar}},\ }\href@noop {}
  {\bibfield  {journal} {\bibinfo  {journal} {ACS Appl. Mater. Interfaces}\
  }\textbf {\bibinfo {volume} {6}},\ \bibinfo {pages} {10917} (\bibinfo {year}
  {2014})}\BibitemShut {NoStop}%
\bibitem [{\citenamefont {Zhang}\ \emph {et~al.}(2004)\citenamefont {Zhang},
  \citenamefont {Martin}, \citenamefont {Friend}, \citenamefont {Schoonen},\
  and\ \citenamefont {Holland}}]{zhang2004mineral}%
  \BibitemOpen
  \bibfield  {author} {\bibinfo {author} {\bibfnamefont {X.~V.}\ \bibnamefont
  {Zhang}}, \bibinfo {author} {\bibfnamefont {S.~T.}\ \bibnamefont {Martin}},
  \bibinfo {author} {\bibfnamefont {C.~M.}\ \bibnamefont {Friend}}, \bibinfo
  {author} {\bibfnamefont {M.~A.}\ \bibnamefont {Schoonen}}, \ and\ \bibinfo
  {author} {\bibfnamefont {H.~D.}\ \bibnamefont {Holland}},\ }\href@noop {}
  {\bibfield  {journal} {\bibinfo  {journal} {J. Am. Chem. Soc.}\ }\textbf
  {\bibinfo {volume} {126}},\ \bibinfo {pages} {11247} (\bibinfo {year}
  {2004})}\BibitemShut {NoStop}%
\bibitem [{\citenamefont {Zhang}\ \emph {et~al.}(2012)\citenamefont {Zhang},
  \citenamefont {Zhou}, \citenamefont {Wu}, \citenamefont {Xu},\ and\
  \citenamefont {Lou}}]{zhang2012unusual}%
  \BibitemOpen
  \bibfield  {author} {\bibinfo {author} {\bibfnamefont {L.}~\bibnamefont
  {Zhang}}, \bibinfo {author} {\bibfnamefont {L.}~\bibnamefont {Zhou}},
  \bibinfo {author} {\bibfnamefont {H.~B.}\ \bibnamefont {Wu}}, \bibinfo
  {author} {\bibfnamefont {R.}~\bibnamefont {Xu}}, \ and\ \bibinfo {author}
  {\bibfnamefont {X.~W.}\ \bibnamefont {Lou}},\ }\href@noop {} {\bibfield
  {journal} {\bibinfo  {journal} {Angew. Chem. Int. Ed.}\ }\textbf {\bibinfo
  {volume} {51}},\ \bibinfo {pages} {7267} (\bibinfo {year}
  {2012})}\BibitemShut {NoStop}%
\bibitem [{\citenamefont {Tang}\ \emph {et~al.}(2015)\citenamefont {Tang},
  \citenamefont {Chen},\ and\ \citenamefont {Yu}}]{tang2015morphology}%
  \BibitemOpen
  \bibfield  {author} {\bibinfo {author} {\bibfnamefont {Y.}~\bibnamefont
  {Tang}}, \bibinfo {author} {\bibfnamefont {T.}~\bibnamefont {Chen}}, \ and\
  \bibinfo {author} {\bibfnamefont {S.}~\bibnamefont {Yu}},\ }\href@noop {}
  {\bibfield  {journal} {\bibinfo  {journal} {Chem. Commun.}\ }\textbf
  {\bibinfo {volume} {51}},\ \bibinfo {pages} {9018} (\bibinfo {year}
  {2015})}\BibitemShut {NoStop}%
\bibitem [{\citenamefont {Corliss}\ \emph {et~al.}(1956)\citenamefont
  {Corliss}, \citenamefont {Elliott},\ and\ \citenamefont
  {Hastings}}]{hastings1956magnetic}%
  \BibitemOpen
  \bibfield  {author} {\bibinfo {author} {\bibfnamefont {L.}~\bibnamefont
  {Corliss}}, \bibinfo {author} {\bibfnamefont {N.}~\bibnamefont {Elliott}}, \
  and\ \bibinfo {author} {\bibfnamefont {J.}~\bibnamefont {Hastings}},\
  }\href@noop {} {\bibfield  {journal} {\bibinfo  {journal} {Phys. Rev.}\
  }\textbf {\bibinfo {volume} {104}},\ \bibinfo {pages} {924} (\bibinfo {year}
  {1956})}\BibitemShut {NoStop}%
\bibitem [{\citenamefont {Lines}\ and\ \citenamefont
  {Jones}(1966)}]{linesMnS1966antiferromagnetism}%
  \BibitemOpen
  \bibfield  {author} {\bibinfo {author} {\bibfnamefont {M.}~\bibnamefont
  {Lines}}\ and\ \bibinfo {author} {\bibfnamefont {E.}~\bibnamefont {Jones}},\
  }\href@noop {} {\bibfield  {journal} {\bibinfo  {journal} {Phys. Rev.}\
  }\textbf {\bibinfo {volume} {141}},\ \bibinfo {pages} {525} (\bibinfo {year}
  {1966})}\BibitemShut {NoStop}%
\bibitem [{\citenamefont {Morosin}(1970)}]{morosin1970striction}%
  \BibitemOpen
  \bibfield  {author} {\bibinfo {author} {\bibfnamefont {B.}~\bibnamefont
  {Morosin}},\ }\href@noop {} {\bibfield  {journal} {\bibinfo  {journal} {Phys.
  Rev. B}\ }\textbf {\bibinfo {volume} {1}},\ \bibinfo {pages} {236} (\bibinfo
  {year} {1970})}\BibitemShut {NoStop}%
\bibitem [{\citenamefont {Danielian}\ and\ \citenamefont
  {Stevens}(1961)}]{danielian1961exchange}%
  \BibitemOpen
  \bibfield  {author} {\bibinfo {author} {\bibfnamefont {A.}~\bibnamefont
  {Danielian}}\ and\ \bibinfo {author} {\bibfnamefont {K.}~\bibnamefont
  {Stevens}},\ }\href@noop {} {\bibfield  {journal} {\bibinfo  {journal} {Proc.
  Phys. Soc. (1958-1967)}\ }\textbf {\bibinfo {volume} {77}},\ \bibinfo {pages}
  {124} (\bibinfo {year} {1961})}\BibitemShut {NoStop}%
\bibitem [{\citenamefont {Trimarchi}\ \emph {et~al.}(2018)\citenamefont
  {Trimarchi}, \citenamefont {Wang},\ and\ \citenamefont
  {Zunger}}]{zunger2018polymorphous}%
  \BibitemOpen
  \bibfield  {author} {\bibinfo {author} {\bibfnamefont {G.}~\bibnamefont
  {Trimarchi}}, \bibinfo {author} {\bibfnamefont {Z.}~\bibnamefont {Wang}}, \
  and\ \bibinfo {author} {\bibfnamefont {A.}~\bibnamefont {Zunger}},\
  }\href@noop {} {\bibfield  {journal} {\bibinfo  {journal} {Phys. Rev. B}\
  }\textbf {\bibinfo {volume} {97}},\ \bibinfo {pages} {035107} (\bibinfo
  {year} {2018})}\BibitemShut {NoStop}%
\bibitem [{\citenamefont {Hohenberg}\ and\ \citenamefont {Kohn}(1964)}]{kohn1}%
  \BibitemOpen
  \bibfield  {author} {\bibinfo {author} {\bibfnamefont {P.}~\bibnamefont
  {Hohenberg}}\ and\ \bibinfo {author} {\bibfnamefont {W.}~\bibnamefont
  {Kohn}},\ }\href@noop {} {\bibfield  {journal} {\bibinfo  {journal} {Phys.
  Rev.}\ }\textbf {\bibinfo {volume} {136}},\ \bibinfo {pages} {B864} (\bibinfo
  {year} {1964})}\BibitemShut {NoStop}%
\bibitem [{\citenamefont {Kohn}\ and\ \citenamefont {Sham}(1965)}]{kohn2}%
  \BibitemOpen
  \bibfield  {author} {\bibinfo {author} {\bibfnamefont {W.}~\bibnamefont
  {Kohn}}\ and\ \bibinfo {author} {\bibfnamefont {L.~J.}\ \bibnamefont
  {Sham}},\ }\href@noop {} {\bibfield  {journal} {\bibinfo  {journal} {Phys.
  Rev.}\ }\textbf {\bibinfo {volume} {140}},\ \bibinfo {pages} {A1133}
  (\bibinfo {year} {1965})}\BibitemShut {NoStop}%
\bibitem [{\citenamefont {Oguchi}\ \emph {et~al.}(1983)\citenamefont {Oguchi},
  \citenamefont {Terakura},\ and\ \citenamefont {Williams}}]{oguchi1983band}%
  \BibitemOpen
  \bibfield  {author} {\bibinfo {author} {\bibfnamefont {T.}~\bibnamefont
  {Oguchi}}, \bibinfo {author} {\bibfnamefont {K.}~\bibnamefont {Terakura}}, \
  and\ \bibinfo {author} {\bibfnamefont {A.~R.}\ \bibnamefont {Williams}},\
  }\href@noop {} {\bibfield  {journal} {\bibinfo  {journal} {Phys. Rev. B}\
  }\textbf {\bibinfo {volume} {28}},\ \bibinfo {pages} {6443} (\bibinfo {year}
  {1983})}\BibitemShut {NoStop}%
\bibitem [{\citenamefont {Raybaud}\ \emph
  {et~al.}(1997{\natexlab{a}})\citenamefont {Raybaud}, \citenamefont {Kresse},
  \citenamefont {Hafner},\ and\ \citenamefont {Toulhoat}}]{raybaud1997ab}%
  \BibitemOpen
  \bibfield  {author} {\bibinfo {author} {\bibfnamefont {P.}~\bibnamefont
  {Raybaud}}, \bibinfo {author} {\bibfnamefont {G.}~\bibnamefont {Kresse}},
  \bibinfo {author} {\bibfnamefont {J.}~\bibnamefont {Hafner}}, \ and\ \bibinfo
  {author} {\bibfnamefont {H.}~\bibnamefont {Toulhoat}},\ }\href@noop {}
  {\bibfield  {journal} {\bibinfo  {journal} {J. Phys. Condens. Matter}\
  }\textbf {\bibinfo {volume} {9}},\ \bibinfo {pages} {11085} (\bibinfo {year}
  {1997}{\natexlab{a}})}\BibitemShut {NoStop}%
\bibitem [{\citenamefont {Raybaud}\ \emph
  {et~al.}(1997{\natexlab{b}})\citenamefont {Raybaud}, \citenamefont {Hafner},
  \citenamefont {Kresse},\ and\ \citenamefont {Toulhoat}}]{raybaud1997abc}%
  \BibitemOpen
  \bibfield  {author} {\bibinfo {author} {\bibfnamefont {P.}~\bibnamefont
  {Raybaud}}, \bibinfo {author} {\bibfnamefont {J.}~\bibnamefont {Hafner}},
  \bibinfo {author} {\bibfnamefont {G.}~\bibnamefont {Kresse}}, \ and\ \bibinfo
  {author} {\bibfnamefont {H.}~\bibnamefont {Toulhoat}},\ }\href@noop {}
  {\bibfield  {journal} {\bibinfo  {journal} {J. Phys. Condens. Matter}\
  }\textbf {\bibinfo {volume} {9}},\ \bibinfo {pages} {11107} (\bibinfo {year}
  {1997}{\natexlab{b}})}\BibitemShut {NoStop}%
\bibitem [{\citenamefont {Tappero}\ and\ \citenamefont
  {Lichanot}(1998)}]{tappero1998comparative}%
  \BibitemOpen
  \bibfield  {author} {\bibinfo {author} {\bibfnamefont {R.}~\bibnamefont
  {Tappero}}\ and\ \bibinfo {author} {\bibfnamefont {A.}~\bibnamefont
  {Lichanot}},\ }\href@noop {} {\bibfield  {journal} {\bibinfo  {journal}
  {Chem. Phys.}\ }\textbf {\bibinfo {volume} {236}},\ \bibinfo {pages} {97}
  (\bibinfo {year} {1998})}\BibitemShut {NoStop}%
\bibitem [{\citenamefont {Hobbs}\ and\ \citenamefont
  {Hafner}(1999)}]{hobbs1999magnetism}%
  \BibitemOpen
  \bibfield  {author} {\bibinfo {author} {\bibfnamefont {D.}~\bibnamefont
  {Hobbs}}\ and\ \bibinfo {author} {\bibfnamefont {J.}~\bibnamefont {Hafner}},\
  }\href@noop {} {\bibfield  {journal} {\bibinfo  {journal} {J. Phys. Condens.
  Matter}\ }\textbf {\bibinfo {volume} {11}},\ \bibinfo {pages} {8197}
  (\bibinfo {year} {1999})}\BibitemShut {NoStop}%
\bibitem [{\citenamefont {Rohrbach}\ \emph {et~al.}(2003)\citenamefont
  {Rohrbach}, \citenamefont {Hafner},\ and\ \citenamefont
  {Kresse}}]{rohrbach2003electronic}%
  \BibitemOpen
  \bibfield  {author} {\bibinfo {author} {\bibfnamefont {A.}~\bibnamefont
  {Rohrbach}}, \bibinfo {author} {\bibfnamefont {J.}~\bibnamefont {Hafner}}, \
  and\ \bibinfo {author} {\bibfnamefont {G.}~\bibnamefont {Kresse}},\
  }\href@noop {} {\bibfield  {journal} {\bibinfo  {journal} {J. Phys. Condens.
  Matter}\ }\textbf {\bibinfo {volume} {15}},\ \bibinfo {pages} {979} (\bibinfo
  {year} {2003})}\BibitemShut {NoStop}%
\bibitem [{\citenamefont {Fender}\ \emph {et~al.}(1968)\citenamefont {Fender},
  \citenamefont {Jacobson},\ and\ \citenamefont
  {Wedgwood}}]{fender1968covalency}%
  \BibitemOpen
  \bibfield  {author} {\bibinfo {author} {\bibfnamefont {B.}~\bibnamefont
  {Fender}}, \bibinfo {author} {\bibfnamefont {A.}~\bibnamefont {Jacobson}}, \
  and\ \bibinfo {author} {\bibfnamefont {F.}~\bibnamefont {Wedgwood}},\
  }\href@noop {} {\bibfield  {journal} {\bibinfo  {journal} {J. Chem. Phys}\
  }\textbf {\bibinfo {volume} {48}},\ \bibinfo {pages} {990} (\bibinfo {year}
  {1968})}\BibitemShut {NoStop}%
\bibitem [{\citenamefont {Zunger}\ \emph {et~al.}(1990)\citenamefont {Zunger},
  \citenamefont {Wei}, \citenamefont {Ferreira},\ and\ \citenamefont
  {Bernard}}]{zunger1990special}%
  \BibitemOpen
  \bibfield  {author} {\bibinfo {author} {\bibfnamefont {A.}~\bibnamefont
  {Zunger}}, \bibinfo {author} {\bibfnamefont {S.-H.}\ \bibnamefont {Wei}},
  \bibinfo {author} {\bibfnamefont {L.~G.}\ \bibnamefont {Ferreira}}, \ and\
  \bibinfo {author} {\bibfnamefont {J.~E.}\ \bibnamefont {Bernard}},\
  }\href@noop {} {\bibfield  {journal} {\bibinfo  {journal} {Phys. Rev. Lett.}\
  }\textbf {\bibinfo {volume} {65}},\ \bibinfo {pages} {353} (\bibinfo {year}
  {1990})}\BibitemShut {NoStop}%
\bibitem [{\citenamefont {Varignon}\ \emph {et~al.}(2019)\citenamefont
  {Varignon}, \citenamefont {Bibes},\ and\ \citenamefont
  {Zunger}}]{zunger2019origin}%
  \BibitemOpen
  \bibfield  {author} {\bibinfo {author} {\bibfnamefont {J.}~\bibnamefont
  {Varignon}}, \bibinfo {author} {\bibfnamefont {M.}~\bibnamefont {Bibes}}, \
  and\ \bibinfo {author} {\bibfnamefont {A.}~\bibnamefont {Zunger}},\
  }\href@noop {} {\bibfield  {journal} {\bibinfo  {journal} {Nat. Commun.}\
  }\textbf {\bibinfo {volume} {10}},\ \bibinfo {pages} {1} (\bibinfo {year}
  {2019})}\BibitemShut {NoStop}%
\bibitem [{\citenamefont {Kraft}\ and\ \citenamefont
  {Greuling}(1988)}]{kraft1988high}%
  \BibitemOpen
  \bibfield  {author} {\bibinfo {author} {\bibfnamefont {A.}~\bibnamefont
  {Kraft}}\ and\ \bibinfo {author} {\bibfnamefont {B.}~\bibnamefont
  {Greuling}},\ }\href@noop {} {\bibfield  {journal} {\bibinfo  {journal}
  {Cryst. Res. Technol.}\ }\textbf {\bibinfo {volume} {23}},\ \bibinfo {pages}
  {605} (\bibinfo {year} {1988})}\BibitemShut {NoStop}%
\bibitem [{\citenamefont {McCammon}(1991)}]{mccammon1991static}%
  \BibitemOpen
  \bibfield  {author} {\bibinfo {author} {\bibfnamefont {C.}~\bibnamefont
  {McCammon}},\ }\href@noop {} {\bibfield  {journal} {\bibinfo  {journal}
  {Phys. Chem. Miner.}\ }\textbf {\bibinfo {volume} {17}},\ \bibinfo {pages}
  {636} (\bibinfo {year} {1991})}\BibitemShut {NoStop}%
\bibitem [{\citenamefont {Sweeney}\ and\ \citenamefont
  {Heinz}(1993)}]{sweeney1993compression}%
  \BibitemOpen
  \bibfield  {author} {\bibinfo {author} {\bibfnamefont {J.~S.}\ \bibnamefont
  {Sweeney}}\ and\ \bibinfo {author} {\bibfnamefont {D.~L.}\ \bibnamefont
  {Heinz}},\ }\href@noop {} {\bibfield  {journal} {\bibinfo  {journal} {Phys.
  Chem. Miner.}\ }\textbf {\bibinfo {volume} {20}},\ \bibinfo {pages} {63}
  (\bibinfo {year} {1993})}\BibitemShut {NoStop}%
\bibitem [{\citenamefont {Xiao}\ \emph {et~al.}(2015)\citenamefont {Xiao},
  \citenamefont {Yang}, \citenamefont {Zhang}, \citenamefont {Wang},
  \citenamefont {Huang}, \citenamefont {Ding}, \citenamefont {Ma},
  \citenamefont {Zou},\ and\ \citenamefont {Zou}}]{b31}%
  \BibitemOpen
  \bibfield  {author} {\bibinfo {author} {\bibfnamefont {G.}~\bibnamefont
  {Xiao}}, \bibinfo {author} {\bibfnamefont {X.}~\bibnamefont {Yang}}, \bibinfo
  {author} {\bibfnamefont {X.}~\bibnamefont {Zhang}}, \bibinfo {author}
  {\bibfnamefont {K.}~\bibnamefont {Wang}}, \bibinfo {author} {\bibfnamefont
  {X.}~\bibnamefont {Huang}}, \bibinfo {author} {\bibfnamefont
  {Z.}~\bibnamefont {Ding}}, \bibinfo {author} {\bibfnamefont {Y.}~\bibnamefont
  {Ma}}, \bibinfo {author} {\bibfnamefont {G.}~\bibnamefont {Zou}}, \ and\
  \bibinfo {author} {\bibfnamefont {B.}~\bibnamefont {Zou}},\ }\href@noop {}
  {\bibfield  {journal} {\bibinfo  {journal} {J. Am. Chem. Soc.}\ }\textbf
  {\bibinfo {volume} {137}},\ \bibinfo {pages} {10297} (\bibinfo {year}
  {2015})}\BibitemShut {NoStop}%
\bibitem [{\citenamefont {Momma}\ and\ \citenamefont {Izumi}(2008)}]{vesta}%
  \BibitemOpen
  \bibfield  {author} {\bibinfo {author} {\bibfnamefont {K.}~\bibnamefont
  {Momma}}\ and\ \bibinfo {author} {\bibfnamefont {F.}~\bibnamefont {Izumi}},\
  }\href@noop {} {\bibfield  {journal} {\bibinfo  {journal} {J. Appl.
  Crystallogr.}\ }\textbf {\bibinfo {volume} {41}},\ \bibinfo {pages} {653}
  (\bibinfo {year} {2008})}\BibitemShut {NoStop}%
\bibitem [{\citenamefont {Bl{\"o}chl}(1994)}]{blochl}%
  \BibitemOpen
  \bibfield  {author} {\bibinfo {author} {\bibfnamefont {P.~E.}\ \bibnamefont
  {Bl{\"o}chl}},\ }\href@noop {} {\bibfield  {journal} {\bibinfo  {journal}
  {Phys. Rev. B}\ }\textbf {\bibinfo {volume} {50}},\ \bibinfo {pages} {17953}
  (\bibinfo {year} {1994})}\BibitemShut {NoStop}%
\bibitem [{\citenamefont {Kresse}\ and\ \citenamefont
  {Furthm{\"u}ller}(1996{\natexlab{a}})}]{kresse1996efficiency}%
  \BibitemOpen
  \bibfield  {author} {\bibinfo {author} {\bibfnamefont {G.}~\bibnamefont
  {Kresse}}\ and\ \bibinfo {author} {\bibfnamefont {J.}~\bibnamefont
  {Furthm{\"u}ller}},\ }\href@noop {} {\bibfield  {journal} {\bibinfo
  {journal} {Comput. Mater. Sci.}\ }\textbf {\bibinfo {volume} {6}},\ \bibinfo
  {pages} {15} (\bibinfo {year} {1996}{\natexlab{a}})}\BibitemShut {NoStop}%
\bibitem [{\citenamefont {Kresse}\ and\ \citenamefont
  {Furthm{\"u}ller}(1996{\natexlab{b}})}]{kresse1996efficient}%
  \BibitemOpen
  \bibfield  {author} {\bibinfo {author} {\bibfnamefont {G.}~\bibnamefont
  {Kresse}}\ and\ \bibinfo {author} {\bibfnamefont {J.}~\bibnamefont
  {Furthm{\"u}ller}},\ }\href@noop {} {\bibfield  {journal} {\bibinfo
  {journal} {Physical review B}\ }\textbf {\bibinfo {volume} {54}},\ \bibinfo
  {pages} {11169} (\bibinfo {year} {1996}{\natexlab{b}})}\BibitemShut {NoStop}%
\bibitem [{\citenamefont {Csonka}\ \emph {et~al.}(2009)\citenamefont {Csonka},
  \citenamefont {Perdew}, \citenamefont {Ruzsinszky}, \citenamefont
  {Philipsen}, \citenamefont {Leb\`egue}, \citenamefont {Paier}, \citenamefont
  {Vydrov},\ and\ \citenamefont {\'Angy\'an}}]{PBEsol}%
  \BibitemOpen
  \bibfield  {author} {\bibinfo {author} {\bibfnamefont {G.~I.}\ \bibnamefont
  {Csonka}}, \bibinfo {author} {\bibfnamefont {J.~P.}\ \bibnamefont {Perdew}},
  \bibinfo {author} {\bibfnamefont {A.}~\bibnamefont {Ruzsinszky}}, \bibinfo
  {author} {\bibfnamefont {P.~H.~T.}\ \bibnamefont {Philipsen}}, \bibinfo
  {author} {\bibfnamefont {S.}~\bibnamefont {Leb\`egue}}, \bibinfo {author}
  {\bibfnamefont {J.}~\bibnamefont {Paier}}, \bibinfo {author} {\bibfnamefont
  {O.~A.}\ \bibnamefont {Vydrov}}, \ and\ \bibinfo {author} {\bibfnamefont
  {J.~G.}\ \bibnamefont {\'Angy\'an}},\ }\href {\doibase
  10.1103/PhysRevB.79.155107} {\bibfield  {journal} {\bibinfo  {journal} {Phys.
  Rev. B}\ }\textbf {\bibinfo {volume} {79}},\ \bibinfo {pages} {155107}
  (\bibinfo {year} {2009})}\BibitemShut {NoStop}%
\bibitem [{\citenamefont {Liechtenstein}\ \emph {et~al.}(1995)\citenamefont
  {Liechtenstein}, \citenamefont {Anisimov},\ and\ \citenamefont
  {Zaanen}}]{AnisimovU}%
  \BibitemOpen
  \bibfield  {author} {\bibinfo {author} {\bibfnamefont {A.~I.}\ \bibnamefont
  {Liechtenstein}}, \bibinfo {author} {\bibfnamefont {V.~I.}\ \bibnamefont
  {Anisimov}}, \ and\ \bibinfo {author} {\bibfnamefont {J.}~\bibnamefont
  {Zaanen}},\ }\href {\doibase 10.1103/PhysRevB.52.R5467} {\bibfield  {journal}
  {\bibinfo  {journal} {Phys. Rev. B}\ }\textbf {\bibinfo {volume} {52}},\
  \bibinfo {pages} {R5467} (\bibinfo {year} {1995})}\BibitemShut {NoStop}%
\bibitem [{\citenamefont {Dudarev}\ \emph {et~al.}(1998)\citenamefont
  {Dudarev}, \citenamefont {Botton}, \citenamefont {Savrasov}, \citenamefont
  {Humphreys},\ and\ \citenamefont {Sutton}}]{uterm}%
  \BibitemOpen
  \bibfield  {author} {\bibinfo {author} {\bibfnamefont {S.~L.}\ \bibnamefont
  {Dudarev}}, \bibinfo {author} {\bibfnamefont {G.~A.}\ \bibnamefont {Botton}},
  \bibinfo {author} {\bibfnamefont {S.~Y.}\ \bibnamefont {Savrasov}}, \bibinfo
  {author} {\bibfnamefont {C.~J.}\ \bibnamefont {Humphreys}}, \ and\ \bibinfo
  {author} {\bibfnamefont {A.~P.}\ \bibnamefont {Sutton}},\ }\href {\doibase
  10.1103/PhysRevB.57.1505} {\bibfield  {journal} {\bibinfo  {journal} {Phys.
  Rev. B}\ }\textbf {\bibinfo {volume} {57}},\ \bibinfo {pages} {1505}
  (\bibinfo {year} {1998})}\BibitemShut {NoStop}%
\bibitem [{\citenamefont {Ceperley}\ and\ \citenamefont
  {Alder}(1980)}]{ceperley1980ground}%
  \BibitemOpen
  \bibfield  {author} {\bibinfo {author} {\bibfnamefont {D.~M.}\ \bibnamefont
  {Ceperley}}\ and\ \bibinfo {author} {\bibfnamefont {B.~J.}\ \bibnamefont
  {Alder}},\ }\href@noop {} {\bibfield  {journal} {\bibinfo  {journal} {Phys.
  Rev. Lett.}\ }\textbf {\bibinfo {volume} {45}},\ \bibinfo {pages} {566}
  (\bibinfo {year} {1980})}\BibitemShut {NoStop}%
\bibitem [{\citenamefont {Perdew}\ \emph {et~al.}(1996)\citenamefont {Perdew},
  \citenamefont {Burke},\ and\ \citenamefont
  {Ernzerhof}}]{perdew1996generalized}%
  \BibitemOpen
  \bibfield  {author} {\bibinfo {author} {\bibfnamefont {J.~P.}\ \bibnamefont
  {Perdew}}, \bibinfo {author} {\bibfnamefont {K.}~\bibnamefont {Burke}}, \
  and\ \bibinfo {author} {\bibfnamefont {M.}~\bibnamefont {Ernzerhof}},\
  }\href@noop {} {\bibfield  {journal} {\bibinfo  {journal} {Phys. Rev. Lett.}\
  }\textbf {\bibinfo {volume} {77}},\ \bibinfo {pages} {3865} (\bibinfo {year}
  {1996})}\BibitemShut {NoStop}%
\bibitem [{See Supplemental Material at URL
  \url{http://will.be.inserted.by.publisher}()}]{SM}%
  \BibitemOpen
  See Supplemental Material at URL \url{http://will.be.inserted.by.publisher},\
  \href@noop {} {}\bibinfo {note} {{} for the files of the generated AFM and PM
  MnS polymorphs' structures used in this work.}\BibitemShut {Stop}%
\bibitem [{\citenamefont {Van De~Walle}\ \emph {et~al.}(2002)\citenamefont {Van
  De~Walle}, \citenamefont {Asta},\ and\ \citenamefont {Ceder}}]{van2002alloy}%
  \BibitemOpen
  \bibfield  {author} {\bibinfo {author} {\bibfnamefont {A.}~\bibnamefont {Van
  De~Walle}}, \bibinfo {author} {\bibfnamefont {M.}~\bibnamefont {Asta}}, \
  and\ \bibinfo {author} {\bibfnamefont {G.}~\bibnamefont {Ceder}},\
  }\href@noop {} {\bibfield  {journal} {\bibinfo  {journal} {CALPHAD}\ }\textbf
  {\bibinfo {volume} {26}},\ \bibinfo {pages} {539} (\bibinfo {year}
  {2002})}\BibitemShut {NoStop}%
\bibitem [{\citenamefont {Allen}\ and\ \citenamefont
  {Watson}(2014)}]{watson2014occupation}%
  \BibitemOpen
  \bibfield  {author} {\bibinfo {author} {\bibfnamefont {J.~P.}\ \bibnamefont
  {Allen}}\ and\ \bibinfo {author} {\bibfnamefont {G.~W.}\ \bibnamefont
  {Watson}},\ }\href@noop {} {\bibfield  {journal} {\bibinfo  {journal} {Phys.
  Chem. Chem. Phys.}\ }\textbf {\bibinfo {volume} {16}},\ \bibinfo {pages}
  {21016} (\bibinfo {year} {2014})}\BibitemShut {NoStop}%
\bibitem [{\citenamefont {Huffman}\ and\ \citenamefont
  {Wild}(1967)}]{huffman1967optical}%
  \BibitemOpen
  \bibfield  {author} {\bibinfo {author} {\bibfnamefont {D.~R.}\ \bibnamefont
  {Huffman}}\ and\ \bibinfo {author} {\bibfnamefont {R.~L.}\ \bibnamefont
  {Wild}},\ }\href@noop {} {\bibfield  {journal} {\bibinfo  {journal} {Physical
  Review}\ }\textbf {\bibinfo {volume} {156}},\ \bibinfo {pages} {989}
  (\bibinfo {year} {1967})}\BibitemShut {NoStop}%
\bibitem [{\citenamefont {Bradley}\ and\ \citenamefont
  {Cracknell}(2009)}]{bradley2009mathematical}%
  \BibitemOpen
  \bibfield  {author} {\bibinfo {author} {\bibfnamefont {C.}~\bibnamefont
  {Bradley}}\ and\ \bibinfo {author} {\bibfnamefont {A.}~\bibnamefont
  {Cracknell}},\ }\href@noop {} {\emph {\bibinfo {title} {The mathematical
  theory of symmetry in solids: representation theory for point groups and
  space groups}}}\ (\bibinfo  {publisher} {Oxford University Press},\ \bibinfo
  {year} {2009})\BibitemShut {NoStop}%
\bibitem [{\citenamefont {Sato}\ \emph {et~al.}(1997)\citenamefont {Sato},
  \citenamefont {Mihara}, \citenamefont {Furuta}, \citenamefont {Tamura},
  \citenamefont {Mimura}, \citenamefont {Happo}, \citenamefont {Taniguchi},\
  and\ \citenamefont {Ueda}}]{sato1997chemical}%
  \BibitemOpen
  \bibfield  {author} {\bibinfo {author} {\bibfnamefont {H.}~\bibnamefont
  {Sato}}, \bibinfo {author} {\bibfnamefont {T.}~\bibnamefont {Mihara}},
  \bibinfo {author} {\bibfnamefont {A.}~\bibnamefont {Furuta}}, \bibinfo
  {author} {\bibfnamefont {M.}~\bibnamefont {Tamura}}, \bibinfo {author}
  {\bibfnamefont {K.}~\bibnamefont {Mimura}}, \bibinfo {author} {\bibfnamefont
  {N.}~\bibnamefont {Happo}}, \bibinfo {author} {\bibfnamefont
  {M.}~\bibnamefont {Taniguchi}}, \ and\ \bibinfo {author} {\bibfnamefont
  {Y.}~\bibnamefont {Ueda}},\ }\href@noop {} {\bibfield  {journal} {\bibinfo
  {journal} {Physical Review B}\ }\textbf {\bibinfo {volume} {56}},\ \bibinfo
  {pages} {7222} (\bibinfo {year} {1997})}\BibitemShut {NoStop}%
\bibitem [{\citenamefont {Orgassa}\ \emph {et~al.}(1999)\citenamefont
  {Orgassa}, \citenamefont {Fujiwara}, \citenamefont {Schulthess},\ and\
  \citenamefont {Butler}}]{orgassa1999first}%
  \BibitemOpen
  \bibfield  {author} {\bibinfo {author} {\bibfnamefont {D.}~\bibnamefont
  {Orgassa}}, \bibinfo {author} {\bibfnamefont {H.}~\bibnamefont {Fujiwara}},
  \bibinfo {author} {\bibfnamefont {T.~C.}\ \bibnamefont {Schulthess}}, \ and\
  \bibinfo {author} {\bibfnamefont {W.~H.}\ \bibnamefont {Butler}},\
  }\href@noop {} {\bibfield  {journal} {\bibinfo  {journal} {Phys. Rev. B}\
  }\textbf {\bibinfo {volume} {60}},\ \bibinfo {pages} {13237} (\bibinfo {year}
  {1999})}\BibitemShut {NoStop}%
\bibitem [{\citenamefont {Hubbard}(1963)}]{hubbard1963electron}%
  \BibitemOpen
  \bibfield  {author} {\bibinfo {author} {\bibfnamefont {J.}~\bibnamefont
  {Hubbard}},\ }\href@noop {} {\bibfield  {journal} {\bibinfo  {journal} {Proc.
  Math. Phys. Eng. Sci.}\ }\textbf {\bibinfo {volume} {276}},\ \bibinfo {pages}
  {238} (\bibinfo {year} {1963})}\BibitemShut {NoStop}%
\bibitem [{\citenamefont {Bruus}\ and\ \citenamefont
  {Flensberg}(2004)}]{bruus2004many}%
  \BibitemOpen
  \bibfield  {author} {\bibinfo {author} {\bibfnamefont {H.}~\bibnamefont
  {Bruus}}\ and\ \bibinfo {author} {\bibfnamefont {K.}~\bibnamefont
  {Flensberg}},\ }\href@noop {} {\emph {\bibinfo {title} {Many-body quantum
  theory in condensed matter physics: an introduction}}}\ (\bibinfo
  {publisher} {OUP Oxford},\ \bibinfo {year} {2004})\BibitemShut {NoStop}%
\bibitem [{\citenamefont {Slater}\ and\ \citenamefont
  {Koster}(1954)}]{slater1954simplified}%
  \BibitemOpen
  \bibfield  {author} {\bibinfo {author} {\bibfnamefont {J.~C.}\ \bibnamefont
  {Slater}}\ and\ \bibinfo {author} {\bibfnamefont {G.~F.}\ \bibnamefont
  {Koster}},\ }\href@noop {} {\bibfield  {journal} {\bibinfo  {journal} {Phys.
  Rev.}\ }\textbf {\bibinfo {volume} {94}},\ \bibinfo {pages} {1498} (\bibinfo
  {year} {1954})}\BibitemShut {NoStop}%
\end{thebibliography}%

\clearpage
\end{document}